\documentclass{article}

\usepackage{arxiv}

\usepackage[utf8]{inputenc} 
\usepackage[T1]{fontenc}    
\usepackage{hyperref}       
\usepackage{url}            
\usepackage{booktabs}       
\usepackage{amsfonts}       
\usepackage{nicefrac}       
\usepackage{microtype}      
\usepackage{lipsum}
\usepackage{graphicx}
\usepackage{color}
\usepackage{natbib}
\usepackage[most]{tcolorbox}
\usepackage{titlesec}
\usepackage{enumitem}
\usepackage{geometry}
\usepackage{tabularx}
\usepackage{multirow}
\usepackage{array}
\usepackage{makecell}
\usepackage{wrapfig}

\usepackage{longtable, array, ragged2e}
\usepackage[table]{xcolor}

\usepackage{threeparttable}

\definecolor{linkblue}{RGB}{40, 70, 120}   
\definecolor{citegreen}{RGB}{50, 100, 70}  
\definecolor{urlgray}{RGB}{80, 80, 110}    

\hypersetup{
  colorlinks=true,      
  linkcolor=linkblue,   
  citecolor=citegreen,  
  urlcolor=urlgray,     
  filecolor=linkblue    
}

\graphicspath{ {./images/} }

\newcommand{\eg}[2]{
    \begin{tcolorbox}[colback=black!5!white,colframe=black,title={#1},breakable, enhanced]
        #2
    \end{tcolorbox}
}

\definecolor{promptsysborder}{HTML}{4A6FA5}
\definecolor{promptuserborder}{HTML}{6B8E5A}
\definecolor{promptasstborder}{HTML}{9B6B9E}
\definecolor{promptbg}{HTML}{F7F7F7}
\definecolor{promptvarcolor}{HTML}{B25400}
\definecolor{promptcfgcolor}{HTML}{006D5B}
\newcommand{\promptvar}[1]{\textcolor{promptvarcolor}{#1}}

\newcommand{\promptmsg}[3]{%
  \par\noindent
  \fcolorbox{#1}{promptbg}{\parbox{\dimexpr\linewidth-2\fboxsep\relax}{%
    {\footnotesize\textbf{\textcolor{#1}{#2}}}\\[2pt]%
    \footnotesize\ttfamily\raggedright #3}}%
  \par\smallskip
}

\title{Total Simulated Survey Error: Designing and Diagnosing Survey Responses from Large Language Models}

\author{
    \textbf{Indira Sen\textsuperscript{1}}\thanks{Corresponding Author: \texttt{indira.sen@uni-mannheim.de}},
    \textbf{Georg Ahnert\textsuperscript{1}},
    \textbf{Leah von der Heyde\textsuperscript{2}},
    \textbf{Jana Lasser\textsuperscript{3}},
    \textbf{Bernd Weiß\textsuperscript{4}}, and \textbf{Markus Strohmaier\textsuperscript{1,2,5}}\\
    \textsuperscript{1}University of Mannheim,
    \textsuperscript{2}GESIS - Leibniz Institute for the Social Sciences,\\
    \textsuperscript{3}University of Graz,
     \textsuperscript{4}University of Duisburg-Essen, \textsuperscript{5}Complexity Science Hub\\
     }

\begin{document}
\maketitle
\begin{abstract}

Large Language models (LLMs), having been trained on vast amounts of human-generated data, may encode the opinions, attitudes, and behaviors of these humans. As such, LLMs show promise in mimicking human-like patterns that facilitate their use in simulating people in a wide variety of contexts. One such context is using LLMs as `silicon samples', i.e., proxies of people in answering survey questions to establish public opinion, design policies, or use as (social) scientific data. However, several critical questions of social biases, generalization, and technical limitations remain, further complicated by a vast design space open to simulation designers. In addition, LLM-simulated surveys often lean on existing survey infrastructure, e.g., survey questions and samples. These artifacts might not be suitable for LLMs or themselves might have errors that are then inherited by the LLM-powered survey simulations. As a result, we lack a systematic understanding of the design space of LLM-generated surveys as well as how these decisions interplay with inherent LLM limitations. In this article, we aim to \textbf{systematically identify, trace, and document limitations in LLM-generated survey responses.} Building on traditions in the quantitative social sciences, specifically survey methodology and measurement theory, we investigate threats to the validity of LLM-generated survey responses
. To do so, we design a framework that enumerates conceptual errors and systematic biases that can occur at different stages of the survey simulation lifecycle. \textbf{Our framework - called the Total Simulated Survey Error (TS2E) Framework - provides a unified and end-to-end perspective on errors in LLM-generated survey data.} We illustrate the framework through a theoretical and empirical case study, which enables survey simulation designers to systematically understand, identify and reflect on errors in LLM-generated surveys. 
\end{abstract}

\section{Introduction}

Computational simulation has a long history in the social sciences as a tool for theory-building and for exploring social processes that are difficult to observe or manipulate directly. Common approaches for social simulations include agent-based models~\citep{helbing2012agent}, microsimulations~\citep{li2013survey}, and system dynamics~\citep{bala2017system}. Furthermore, simulation approaches are also frequently used to address missing data issues, e.g., in survey research~\citep{kalton1982imputing, little2019, littleMissingDataAnalysis2024}. 
However, while simulations have many potentials for advancing social science, inflexible rule-based simulations and statistical models have prevented generalizable inferences. The advent of large language models (LLMs) has renewed interest in generalizable social simulations~\citep{larooij2025large}. Trained on vast corpora of human-generated text, LLMs encode rich, if imperfect, representations of human attitudes, knowledge, and behavioral tendencies, and have demonstrated an ability to produce human-like outputs across many tasks. This capacity has prompted burgeoning uses of LLMs as proxies for humans in social and behavioral research; the use of these LLM-powered human proxies has been dubbed as ``\textit{in-silico} samples''~\citep{argyle2023out}, ``digital twins''~\citep{peng2025mega}, or ``Homo silicus''~\citep{horton2023large}. 

We specifically zoom in on one type of \textit{in-silico} sampling, \emph{\textbf{LLM-generated survey responses}}: the use of LLMs to produce survey responses that would otherwise be solicited from human respondents.\footnote{For brevity, we interchangeably use `LLM-generated surveys', `LLM-generated survey responses', and `LLM-simulated responses.' While LLMs can also be used for other parts of the survey lifecycle, e.g., to generate survey questions fielded to human participants~\citep{rothschild2026responsible,vonderheyde2026AILoopSystematic}, these are outside the scope of our framework.} LLM-generated surveys entail using LLMs to simulate the characteristics of a specific individual or group, often through a \textbf{persona}, and answer survey questions on their behalf. The key underlying assumption behind the potential of these simulations is the notion of `algorithmic fidelity'~\citep{argyle2023out} --- that LLMs, having been trained on vast amounts of human-generated data, can be steered to mimic the behavior of diverse human subgroups. However, current empirical findings have surfaced the limits of these simulations~\citep{wang2025large,bisbee2024synthetic,chapekis2026ai}, \textit{inter alia}. These limitations might arise due to many reasons, including gaps and skews in the training data of LLMs~\citep{elazar2024s,gururangan2022whose,lucy2024aboutme}, while post-training processes used for state-of-the-art chatbot LLMs, like ChatGPT, might further impact model behavior~\citep{binz2026post,grief2026emulate,lyman2025balancing}.

Therefore, establishing whether such simulations are error-free is far from settled. Here, \textbf{error} refers to \textit{random or systematic differences between the true value of a construct and the LLM-based measurement of it}. Most validation efforts to date compare LLM-generated responses against a human benchmark using some similarity metric, and report aggregate match between real and simulated data \citep{argyle2023out, Sanders2023Demonstrations, hu2025simbench, santurkar2023whose, 10.1145/3701716.3715591}, \textit{inter alia}. Yet, recent work has shown that high aggregate matches can, nonetheless, coexist with severe distortions in variance, subgroup heterogeneity, and downstream statistical relationships, such as correlations or multivariate effects ~\citep{bisbee2024synthetic, doi:10.1177/08944393251337014, holtdirk_in-context_2026, quPerformanceBiasesLarge2024}. Concurrently, existing work has documented many failure modes of LLM-based simulations---social bias, sycophancy, homogenization, and ``alienness'' \citep{anthis2025llm}---and has begun to quantify overall error~\citep{hu2025simbench,cummins2026threat}. \textbf{However, what is missing is a systematic account of \emph{why} these errors arise and \emph{where} they originate in the simulation pipeline}. In particular, the existing literature does not separate errors that stem from \textbf{inherent limitations of current LLMs} from those introduced by \textbf{a researcher's design choices}. Nor do these evaluations systematically distinguish errors that compromise \emph{what} is measured from those that compromise \emph{whose} behavior is represented \citep{heyde2025who}. Without such a decomposition, simulation designers lack a principled basis to diagnose, attribute, and mitigate error.

We address this evaluation gap by developing a framework that traces the origins of error across the entire lifecycle of an LLM-generated survey. Building on quantitative measurement theory, and specifically the Total Survey Error (TSE) framework \citep{groves2010total}, we adapt the well-established distinction between \emph{measurement} errors (errors in how a construct is defined, operationalized and answered by respondents) and \emph{representation} errors (errors in how a well-defined target population is accessed and approximated) to the specific affordances of current LLM technology. The TSE framework has previously been extended to novel quantitative social science data and methods~\citep{weiss2025conceptualizing,daikeler2025assessing}, e.g., `big data' \citep{amaya2020total}, online behavioral traces \citep{sen2021total}, web tracking~\citep{bosch2022survey}, and data donations \citep{boeschoten2020digital}. The Total Error perspective has also been applied to evaluations of LLMs' general capabilities~\citep{messing2026hidden}. We argue that it provides a natural foundation for systematically reasoning about the quality of LLM-generated survey data as well.

\begin{figure*}
    \centering
    \includegraphics[width=.94\textwidth]{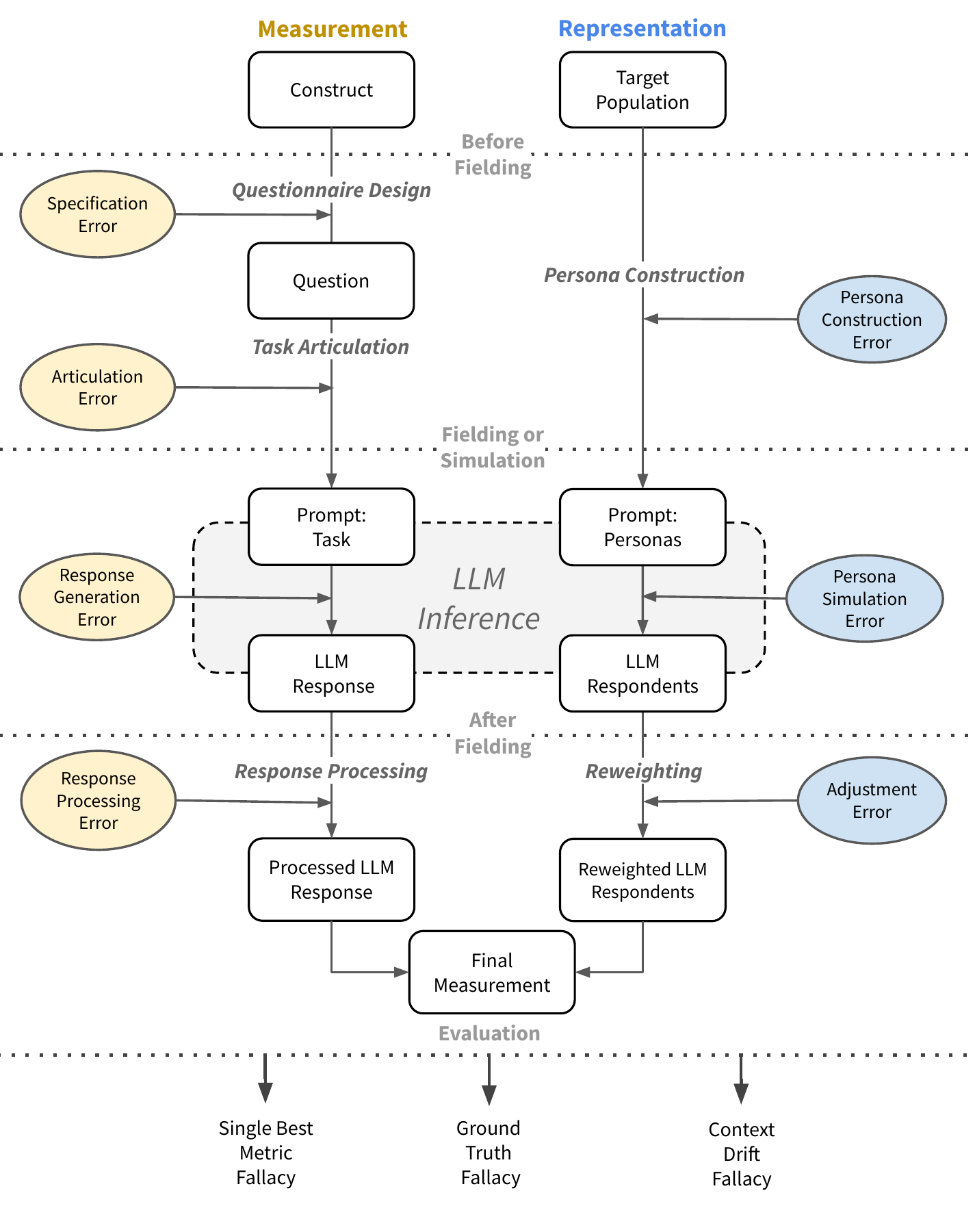}
    \caption{\textbf{The Total Simulated Survey Error (TS2E) Framework: Representation and Measurement Errors, and Evaluation Fallacies.} We show an idealized lifecycle of a simulated survey with different design choices and artifacts. Such simulations can have measurement errors (highlighted in yellow), i.e., errors in \textit{how} an attitude or behavior is simulated, and representation errors (in blue), i.e., errors in \textit{whose} attitudes or behavior is simulated. These errors emerge from inherent limitations in current LLMs (persona simulation errors), the design choices of the simulation designer (e.g., persona construction error), and the interplay between these factors (e.g., response generation error). While the errors can be quantified using reference data (usually human survey data), the three evaluation fallacies describe logical errors in comparing LLM-generated survey data against such reference datasets.}
    \label{fig:ts2e}
\end{figure*}

In this paper, we introduce the \textbf{\emph{Total Simulated Survey Error} (TS2E) framework, a conceptual framework that enumerates the measurement and representation errors that can arise at each step of the LLM-generated survey lifecycle} (Figure~\ref{fig:ts2e} and Section~\ref{sec:ts2e}). The framework distinguishes errors due to design choices of a researcher from those due to inherent LLM limitations, and extends the TSE framework with LLM-specific errors, e.g., the \textbf{persona construction error} due to how the persona of simulated respondents is designed. The framework also introduces \textbf{evaluation fallacies} that might plague the validation of LLM-generated survey responses, especially when they are benchmarked against human responses (Section~\ref{sec:eval_errors}). 
In addition to illustrating these errors with theoretical examples, we provide an empirical case study that applies the framework using a multiverse analysis over simulating vote choice during the 2024 Presidential Election in the U.S. (Section~\ref{sec:case_study}). Finally, based on the framework, we compile a checklist that can guide simulation designers in documenting LLM-generated survey data ( Section~\ref{app:checklist} in the Appendix).

\begin{figure}[h!]
    \centering
    \includegraphics[width=0.99\linewidth]{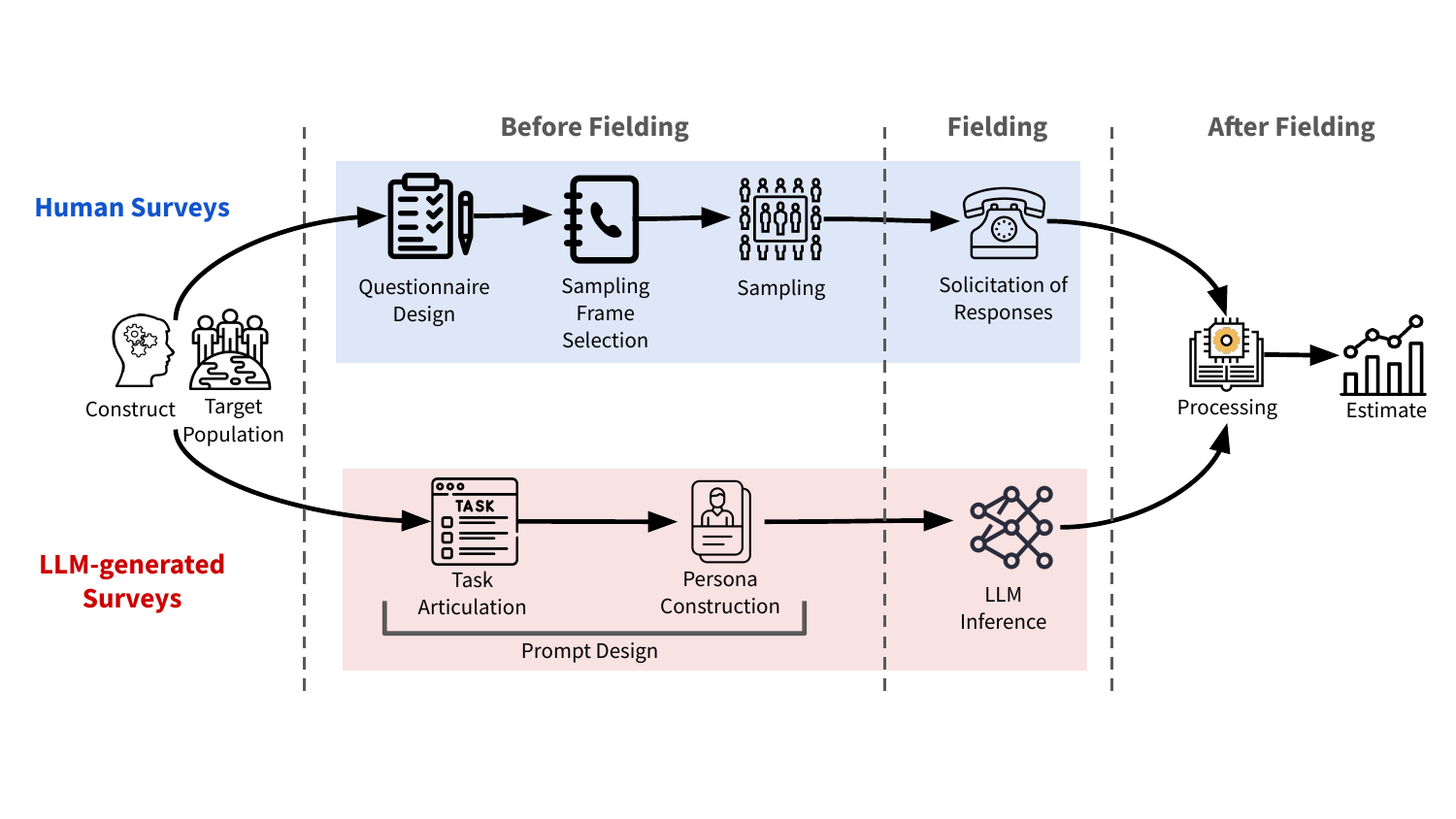}
    \caption{\textbf{Contrasting the Lifecycle of Human and LLM-generated Survey Responses.} For both, the lifecycle has three phases --- (1) \textbf{before fielding}, when the construct is converted to a questionnaire for humans or a task for LLMs and the target population is approximated via a sample of human respondents or personas for LLM respondents, (2) \textbf{fielding}, when responses are obtained from human respondents or via LLM inference, and (3) \textbf{after fielding}, when responses are processed into a final survey estimate.}
    \label{fig:lifecyle}
\end{figure}

Beyond organizing the literature on potentials and pitfalls in LLM-based survey simulations, the TS2E framework is intended as a practical diagnostic tool. By decomposing error by source and by lifecycle stage, it enables simulation designers to isolate which design choices most degrade their estimates, to document errors that lie beyond their control because they are due to inherent LLM limitations, and to reason about trade-offs among competing error sources---much as the TSE framework reframed survey design as an optimization problem over multiple error components rather than a single quality metric. More broadly, we hope the framework offers a common vocabulary facilitating interdisciplinary research in LLM-generated surveys. In that role, the framework could support the development of reporting and evaluation standards and benchmarks for LLM-generated survey data, and help move the field from asking \emph{whether} LLMs can simulate humans toward asking, more precisely, \emph{when}, \emph{for whom}, and \emph{with what errors}.

\section{A Formal Definition of Human and LLM-Generated Survey Responses}\label{sec:definition}

Designing a survey, either fully human-generated or simulated, requires a sequence of decisions allowing the survey designer to estimate a population parameter defined on a theoretical construct. Each unit $i$ of the target population $U=\{1, \ldots, N\}$ has a true value $\mu_i$ of the construct; the target parameter is, e.g., the population mean $\overline{\mu}_U= \frac{1}{N}\sum_{i \in U} \mu_i$, estimated by a survey statistic such as $\bar{y}$. 
An idealized version of this lifecycle shows the different decision steps in Figure~\ref{fig:lifecyle}. These decisions are organized around three phases — (1) before, (2) during, and (3) after fielding of the survey. A summary of notations used in this paper for defining human and LLM survey responses are in Table~\ref{tab:notation}.

\textbf{Before Fielding of the Survey.} The lifecycle begins with defining the objectives that motivate the survey. This means specifying the constructs of interest alongside the target population in which this construct is to be measured. The construct is measured via a survey \textit{question} $q$.\footnote{A construct can also be measured via multiple interrelated questions or items, however, for simplicity, we state that $q$ represents all questions associated with a single construct.} A similar \textbf{questionnaire design} process is followed for both human and LLM-generated surveys. $Y_i$ is $i$'s ideal response to $q$.

Concurrent with questionnaire design, in human-generated surveys, the target population is made accessible through a sampling frame $F$, from which a sample $S$ is drawn.

For LLM-generated surveys, the target population can be operationalized in various ways, but usually relies on \textbf{`personas'} of varying depth~\citep{gordon2026richer}; on the one hand, fine-grained personas can be used to mimic the attributes of individual survey respondents~\citep{argyle2023out,ahnert-etal-2026-survey}. In these cases, information about respondents is repurposed from existing human-generated surveys to create \textbf{`digital twins'}: \citet{argyle2023out}, for instance, use the respondent pool of the American National Election Studies (ANES) as a source for creating personas to simulate the U.S. electorate. On the other hand, more coarse approaches involve prompting the LLM to simulate responses of the whole population or coarse subgroups of the target population via \textbf{`digital populations'}. The latter was done by~\citet{hu2025simbench}, where the persona refers to the entire target population, e.g., ``You are an American.'' or by~\citet{santurkar2023whose}, where the persona refers to a single demographic subgroup, e.g. ``You are a Democrat.''\footnote{For population-level personas, the LLM can still be instructed to simulate an individual person, where the individual belongs to the population or population subgroup.}

\begin{figure}
    \centering
    \includegraphics[width=0.99\linewidth]{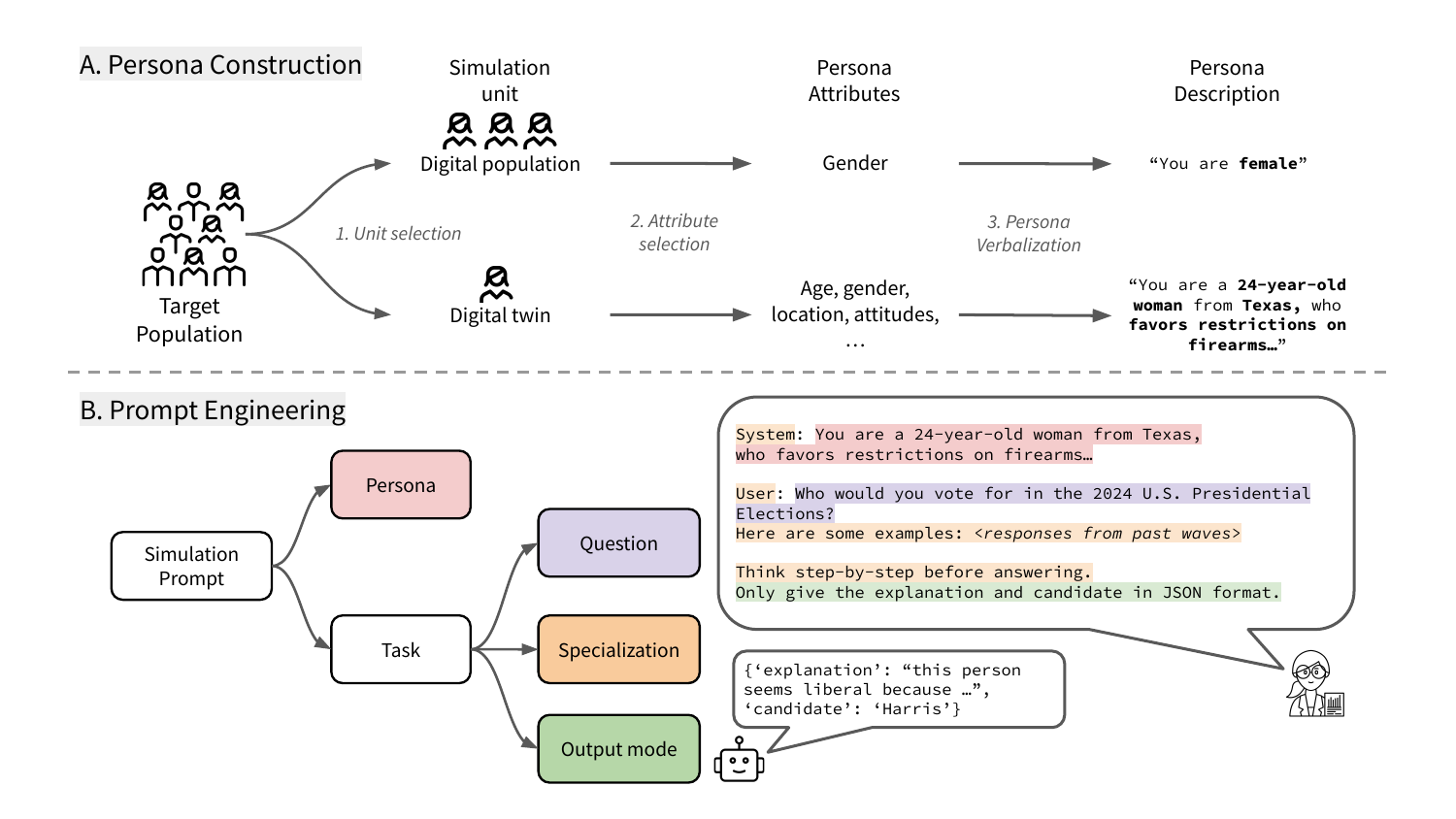}
    \caption{\textbf{A) Persona Construction and B) Prompt Engineering for LLM-simulated Surveys.} In panel (A), we demonstrate how units of the target population are converted into \textbf{simulation units}, based on some auxiliary data source that grounds the simulation, e.g., existing survey datasets. Units of the target population are transformed into simulation units. These simulation units are further transformed into \textbf{persona prompts} based on specific persona attributes. In panel (B), we see how a simulation prompt decomposes into two top-level categories: (1) the \textbf{Persona}, and (2) the \textbf{Task} an LLM is supposed to fulfill while embodying a persona. The latter consists of the survey question the LLM is supposed to answer (\textbf{Question}), with what guidance (\textbf{Specialization}), and how it is supposed to answer (\textbf{Output}). The bottom right panel shows an example of all of these elements, where an LLM is prompted to roleplay as an American individual with specific attitudinal characteristics and asked to declare their vote choice in JSON format. The LLM is specialized by using the system prompt for the persona, few-shot examples of past survey responses, and chain-of-thought reasoning. Suboptimal choices in prompt engineering can lead to misrepresentation of the entities being simulated (representation errors) or invalid measurements of the construct (measurement errors).}
    \label{fig:prompt}
\end{figure}

Therefore, the target population $U$ is mapped to a set of \textbf{simulation units} $V$. A simulation unit is denoted as $j \in V$. We term the process of representing units of the target population through personas as \textbf{persona construction}; it entails several steps --- selecting an appropriate simulation unit,\footnote{Simulation units are the units of observation in LLM-generated surveys. We use the term interchangeably with `simulated units' or `simulated respondents.'} selecting relevant attributes of these units, and finally generating textual personas with cues signaling these attributes (c.f Figure~\ref{fig:prompt}(A)). 

\begin{table*}[t]
\centering
\small
\renewcommand{\arraystretch}{1.25}
\begin{tabularx}{\textwidth}{@{}l X l X@{}}
\toprule
\textbf{Symbol} & \textbf{Meaning} & \textbf{Symbol} & \textbf{Meaning} \\
\midrule
$U$ & Target population & $V$ & LLM simulation units, $j\in V$ \\
$i$ & A single unit of the target population, i.e., a potential human survey respondent & $j$ & A single LLM-simulated unit \\
$\mu_i$ & True value of $\mu$ for unit $i$ & $f_{persona}$ & Persona construction function \\
$q$ & Survey question & $p_j$ & Persona prompt of $j$ \\
$Y_i$ & Ideal response of $i$ to $q$ & $t$ & Task given to the LLM from the adapted question \\
$F$ & Sampling frame & $f_{LLM}$ & LLM response process \\
$S$ & Sample drawn from $F$ & $y^{sim}_j$ & Simulated response of $j$ to $t$, steered by persona $p_j$ \\
$R$ & Human respondents, $R\subseteq S\subseteq F\subseteq U$ & $f_{proc}$ & Response processing (for both humans and LLMs) \\
$f_h$ & Human response process & $f_{adj}$ & Adjustment (e.g., reweighting, for both) \\
$y^h_i$ & $i$'s response to $q$ & $\bar{y}_h,\ \bar{y}_{sim}$ & Final statistic (human, simulated) \\
\bottomrule
\end{tabularx}
\caption{Notation used in this paper to define human and LLM-generated survey responses.}
\label{tab:notation}
\end{table*}

Each simulation unit $j \in V$ represents a set of population units $U_j \subseteq U$.
Personas are \emph{individual-level} (akin to digital twins) if $U_j = \{i\}$ for a single unit $i$, and
\emph{group-level} (akin to digital populations) if $|U_j| > 1$,  e.g., $U_j = \{i \in U : \text{party}(i) = \text{Democrat}\}$,
or $U_j = U$ when a single persona represents the whole subgroup or population. Within a design,
the sets $U_j$ do not overlap, although each $U_j$ may be defined by
multiple, intersecting group memberships (e.g., Black women who are Democrats). Each simulated unit is described by a
\emph{persona prompt}:
\[
  p_j = f_{persona}(j),
\]
obtained from a persona construction function $f_{persona}(\cdot)$ that defines the
characteristics of the units in $U_j$ (for a group, those shared by its members). This process is illustrated in Figure~\ref{fig:prompt}(A). Therefore, when existing survey datasets are used to create individual personas, each respondent's characteristics are used to create one persona per respondent, e.g., ~\citep{ahnert-etal-2026-survey,rupprecht2025german}; on the other hand, for more coarse personas, multiple respondents sharing a characteristic are mapped to the same single persona, e.g., ~\citep{santurkar2023whose,hu2025simbench}.

For LLM-generated surveys, we could prompt LLMs with the survey question and persona. However, models often require further steering or specialization to obtain responses in the desired format. We term this process \textbf{task articulation}, where $q$ is adapted to a version suitable for the LLM into $t$. As shown in Figure~\ref{fig:prompt}(B), a prompt is a structured artifact with distinct components: the (1) \textbf{persona} of the prompt, which includes the persona attributes the LLM should embody for the chosen simulation units, and (2) the simulation \textbf{task} the LLM is prompted to fulfill. The task further contains several facets:

\begin{itemize}
    \item the \textbf{question} the persona-steered LLM should answer; 
    \item the \textbf{specializations} that augment model capabilities, such as few-shot examples, background knowledge, using the system prompt, or reasoning via an inherent reasoning model or through chain-of-thought instructions;
    \item and the \textbf{output modality} that instructs the LLM to answer in a certain format, e.g., open-text~\citep{krsteski2026valid}, JSON~\citep{ahnert-etal-2026-survey}, with distributions~\citep{meister_benchmarking_2025}, or one single answer option for multiple-choice questions~\citep{santurkar2023whose}. 
\end{itemize} 

\textbf{Fielding of the Survey.} In human-generated surveys, the sampled individuals receive the survey questionnaire and choose to participate in the survey. Their answers are recorded. These are the survey's \emph{respondents}  $R$, where $R \subseteq S \subseteq F \subseteq U$. 

Each sampled human respondent $i \in R$
produces an observed \emph{response}
\[
y_i^{h} = f_h(q, i), \qquad i \in R,
\]

where $f_h(\cdot)$ denotes the unobserved, partly stochastic human response process that maps a
questionnaire and a respondent's true attitudes or behavior to the expressed response.

For LLM-simulated surveys, fielding the survey proceeds via \textbf{LLM inference}, i.e., the LLM is prompted to respond. The designer selects and configures the LLM backend, including model architecture, size, alignment type (base vs. instruction-tuned), and inference hyperparameters, such as temperature or restrictions of how the LLM should answer. 

Thus, an LLM respondent $j$ is initialized with a prompt designed in the earlier step consisting of the task $t$ and persona $p_j$. The LLM then generates outputs, ideally based on the response format specified in the task, via LLM inference. Therefore, we obtain an LLM-simulated response defined by:
\[
y_j^{sim} = f_{LLM}\big(t, p_j\big), \qquad j \in V,
\]

where $f_{LLM}(.)$ is also a stochastic function mapping of the task and persona to the chosen LLM's predicted response. A simulated survey response is specified by $(t, p_j, f_{LLM})$, while a human survey
response is specified by $(q, i, f_h)$. Both start with the same question $q$ and target population $U$,
but the simulated version replaces (1) the question $q$ with task $t$, (2) the respondent $i$ with a persona $p_j$, and (3) the human response generation function $f_h$ with a LLM-response generation function $f_{LLM}$. Gaps between the human responses ($y_i^h$) and LLM responses ($y_j^{sim}$), or \textit{errors} in the simulation, can then be traced to the task articulation ($q$ vs.\ $t$), to persona construction ($f_{persona}$), or to the LLM's response generation and persona simulation ($f_{LLM}(t, p)$). 

\textbf{After Fielding of the Survey.} In the final phase, human and simulated survey responses can be treated almost identically. The unprocessed survey responses, either human or simulated, are converted into a form suitable for analysis using a \textbf{response processing} approach signified by $f_{proc}$. $f_{proc}$ might also involve using LLMs, e.g., using LLM for coding or cleaning survey responses~\citep{mellon2024ais}. The survey designer might also apply \textbf{post-hoc adjustments}, such as reweighting of responses and respondents to better match the target population using an adjustment technique, $f_{adj}$. After processing and adjusting responses, the survey designer conducts the substantive analysis to obtain the final survey statistic $\bar{y}_h$ for human-generated surveys and $\bar{y}_{sim}$ for LLM-generated surveys. 

Therefore:
\[
\bar{y}_h = \frac{1}{|R|} \sum^{|R|} f_{adj}(f_{proc}(y_i^{h})), \qquad i \in R,
\]

and 

\[
\bar{y}_{sim} = \frac{1}{|V|} \sum^{|V|} f_{adj}(f_{proc}(y_j^{sim})), \qquad j \in V,
\]

\begin{table*}[t]
\centering
\small
\renewcommand{\arraystretch}{1.3}
\newcolumntype{Y}{>{\raggedright\arraybackslash}X}
\begin{tabularx}{\textwidth}{@{}>{\raggedright\arraybackslash}p{1.9cm} Y >{\raggedright\arraybackslash}p{2.4cm} Y >{\raggedright\arraybackslash}p{2.9cm}@{}}
\toprule
 & \textbf{Human-generated survey} & \textbf{Human survey error (TSE)} & \textbf{LLM-generated survey} & \textbf{LLM survey error (TS2E)} \\
\midrule
\multicolumn{5}{c}{\textit{Before fielding}} \\
\midrule
\textbf{Target population} & $U$, with units $i\in U$ & --- & Same $U$, represented by simulation units $j\in V$ & --- \\
\textbf{Instrument} & Survey question $q$ & Specification (M) & Task $t$: the $q$ adapted for an LLM, plus output modality and specialization & Specification (M); Articulation (M) \\
\textbf{Who is surveyed} & Sample $S$ drawn from frame $F$ & Coverage (R); sampling (R) & Persona prompt $p_j=f_{persona}(j)$ for each simulation unit $j$ & Persona construction (R) \\
\midrule
\multicolumn{5}{c}{\textit{Fielding}} \\
\midrule
\textbf{Respondents} & $R\subseteq S\subseteq F\subseteq U$, the sampled units who respond & Nonresponse (R) & All simulation units $j\in V$ & --- \\
\textbf{Survey response} & $y^h_i=f_h(q,i)$, for $i\in R$ & Response (M) & $y^{sim}_j=f_{LLM}(t,p_j)$, for $j\in V$ & Response generation (M); Persona simulation (R) \\
\midrule
\multicolumn{5}{c}{\textit{After fielding}} \\
\midrule
\textbf{Processing} & $f_{proc}$ applied to each $y^h_i$ & Processing (M) & $f_{proc}$ applied to each $y^{sim}_j$ & Response processing (M) \\
\textbf{Adjustment} & $f_{adj}$ & Adjustment (R) & $f_{adj}$ & Adjustment (R) \\
\textbf{Final statistic} & $\bar{y}_h=\frac{1}{|R|} \sum^{|R|} f_{adj}(f_{proc}(y_i^{h})),$ $i \in R$ & Total survey error & $\bar{y}_{sim} = \frac{1}{|V|} \sum^{|V|} f_{adj}(f_{proc}(y_j^{sim})),$ $j \in V$ & Total simulated survey error \\
\bottomrule
\end{tabularx}
\caption{Human-generated vs.\ LLM-generated surveys by lifecycle stage, with the error each step can introduce under TSE and TS2E. (M) = measurement error, (R) = representation error.}
\label{tab:human_vs_llm}
\end{table*}

For LLM-generated survey data, there could be a further \textbf{evaluation} step, where the simulated survey responses are compared against available ground truth using an evaluation metric or against theoretical expectations. The choice of simulation unit and its mapping to the target population also impacts evaluation; if simulation units map to individuals from a reference dataset, then individual-level performance metrics, such as F1 scores, can be used to gauge simulation performance. 
Individual responses can also be aggregated to create group-level measures, e.g., vote choice of Democrats. Therefore, distribution-level metrics that measure the similarity between response distributions can be applied irrespective of the simulation unit.



\section{The Total Simulated Survey Error (TS2E) Framework}\label{sec:ts2e}

The previous section laid out the design choices the simulation designer must make to set up their simulation. These design choices, as well as the inherent drawbacks of current LLMs, can lead to differences between the true measure and the obtained measure. These differences are called \textbf{errors}; they can be random (variance) or systematic (bias) and they present a threat to the validity and reliability of the simulation. Mirroring the Total Survey Error (TSE) framework \citep{groves2010total}, these errors can be of two types -- \textbf{measurement errors}, which reduce construct validity, and \textbf{representation errors}, which reduce external validity or generalizability (See Appendix Section ~\ref{app:tse} for a primer on the TSE). For LLM-generated surveys, a further threat to quality arises in how they are evaluated; unlike representation and measurement errors that can be quantified by contrasting against a reference set of responses, ideally human responses to the same survey, the evaluation step can entail several logical fallacies in how this reference dataset is selected and compared  against. Taken together, \textbf{all design-driven and LLM-inherent errors, as well as evaluation fallacies, make up the Total Simulated Survey Error (TS2E) Framework}.

In the following subsections, we revisit each stage of an idealized simulated survey lifecycle and introduce the errors associated with each of them (Figure~\ref{fig:ts2e}). Each error is defined as a gap between artifacts obtained at different stages of the lifecycle; therefore, as we progress across the stages of a simulation, each stage involves a design choice and incurs an error. We illustrate these design choices and errors using a case study inspired by recent work on simulating survey responses~\citep{argyle2023out}. In this simulation, we want to simulate the vote choices of American voters in the 2024 presidential election. Therefore, our \textbf{construct} is the voting preferences of our \textbf{target population}, i.e., American voters.

\subsection{Before Fielding of the Survey}

This is the initialization stage of a simulation with the following design choices: 

\textbf{Questionnaire Design. }We design prompts that will guide the LLM respondents, and errors in this step can affect both the measurement and representation dimensions. In translating the construct into a questionnaire, we encounter the first measurement error in LLM-generated surveys -- \emph{\textbf{specification error}}, which is a gap between the construct definition and the question(s) used to operationalize the construct(s).\footnote{Specification error is also referred to as `validity'~\citep{groves2005survey}} In principle, specification errors in surveys are independent of whether the respondents are human or simulated; specification error depends on how well the construct and all of its facets can be translated into survey questions. Therefore, specification error is analogous to its counterpart in the TSE.

\eg{Specification Error}{
\textbf{Definition. }The gap between the construct and the question used to measure it.
\\
\\
\textbf{Example. }For interviewing American voters, the ANES used the following question: \textit{``How about the election for President? Who did you vote for, or did you not vote for President?''}\footnote{\href{https://electionstudies.org/data-tools/anes-variable/variable.html?year=2024&variable=V242068}{electionstudies.org/data-tools/anes-variable/variable.html?year=2024\&variable=V242068}} Vote choice is a relatively straightforward construct to translate into a survey question, unlike sexist or racist attitudes, which might have several subconstructs; More complex vote choice questions include those on voting in the primaries, e.g., \textit{``Did you vote in a Presidential primary election or caucus this year?''}\footnote{\href{https://electionstudies.org/data-tools/anes-variable/variable.html?year=2024&variable=V241031}{electionstudies.org/data-tools/anes-variable/variable.html?year=2024\&variable=V241031}} Accurate answers to such questions hinge on a real or simulated respondent's ability to correctly interpret `caucus' or `primary', which might require a certain degree of political literacy. The lack of accessibility of such questions introduces \textbf{specification error}.  
}

\textbf{Task Articulation. }Questions meant for humans might need to be changed to make them more suitable for LLM-based inference, e.g., instead of providing all questions in a survey battery at once, each prompt might include each separately. Furthermore, to assess the sensitivity and brittleness of LLMs~\citep{sclar2023quantifying}, a simulation designer might automatically or manually paraphrase survey questions and generate variations of the original. However, this adaptation and even minor paraphrases might alter the meaning of the survey question and lead to construct drift, causing an \emph{\textbf{articulation error}}. Articulation also impacts other facets of the survey question besides the actual question itself, e.g., the answer options and their structure, as well as the other aspects of the task, i.e., the output format or the specialization (c.f. Figure~\ref{fig:prompt}(B)). Articulation error is comparable to design-specific response error (also called `measurement error' in~\citet{groves2005survey}) in the TSE.

\eg{Articulation Error}{
\textbf{Definition. }The gap between the question and the task formulated to prompt the LLM. 
\\
\\
\textbf{Example. }For simulating the vote choice of the American populace, one could reuse the exact survey question from the ANES. However, since it's a follow-up from a preceding question (\textit{``How about the election for President? Who did you vote for, or did you not vote for President?''}), it might require further contextualizing when prompting LLMs. Therefore, the question is rephrased to \textit{``Will you vote in the 2024 U.S. presidential election and if so, for whom?''}. Furthermore, the simulation designer might include answer options. However, similar to surveys of humans, if the only answer options included are the major candidates, that might steer the LLM respondents to only choose an answer from those candidates. 
}

\textbf{Persona Construction. }Just as the question is translated into a prompt, units of the target population are converted into simulation units, each of which is represented by a persona. The persona, as well as the task, determine the behavior of the LLM. As we discuss in Section~\ref{sec:definition}, personas may or may not map to individual survey respondents. Therefore, a simulation designer might choose to model the entire target population in a single persona, disaggregate the target population into smaller subgroups, or even down to the level of individuals (Figure~\ref{fig:prompt}(A)). The latter could even entail converting \textit{all} units of the target population into personas; however, this is infeasible for large target populations such as national populations. Therefore, existing survey samples~\citep{argyle2023out,park2024generative,rupprecht2025german} or even census data~\citep{jennings2025nemotron} can be reused to create personas.

As LLM-powered social simulations are a novel research direction, persona prompting lacks standardization w.r.t formatting and content. There are a wide range of persona cues~\citep{sen2025missing}, e.g., to simulate a woman, an LLM might be prompted with different expressions of gender identity (``you are a \textbf{woman}'' vs. ``you are \textbf{female}''), while the persona prompt can have different formats, e.g., second-person direct format: ``You are a woman. Please answer [QUESTION].'' vs. the so-called `QA' or `interview' format: ``What is your gender: woman. What is your answer to [QUESTION]: ''. Another dimension of the persona is its \textbf{composition} --- which attributes make up a persona, e.g., gender, age, attitudes, etc. Personas might also include detailed backstories, either automatically generated~\citep{moon2024virtual} or based on in-depth interviews of real individuals~\citep{park2024generative,gordon2026richer}.\footnote{Personas are one means of steering LLMs to simulate respondents better; other approaches include further training of models (`specialization' in Figure~\ref{fig:prompt}(B)). Personas are nonetheless needed for inference-time steering, i.e., steering an LLM via prompting.}

The \textbf{semantic expression, structure, and composition} of a persona can all impact how well LLMs impersonate the real-world entities that the simulation designer wants to simulate~\citep{lutz2025prompt}. Previous work has shown that spurious persona variables can hinder effective simulation~\citep{de2025principled}. Furthermore, when using human survey infrastructure to create personas, e.g., individual responses from existing survey samples or the census, errors in these infrastructure, such as coverage or sampling error, are inherited by the LLM simulation. We term these representation errors resulting from the simulation designer's simulation unit selection and persona design as \emph{\textbf{persona construction error}}. The composition of a persona is of particular note; it is not clear which attributes should make up a persona and whether this selection should be determined based on an \textit{a priori} theory, constructed in a data-driven manner, or a combination of both. Persona construction error is unique to LLM-generated surveys --- to some extent, it has analogues with both coverage and sampling errors in the TSE since it has to do with the construction of units that are to be surveyed. However, depending on the type of simulation, there might be no explicit sampling involved or the sample might be borrowed from existing survey infrastructure, thereby also inheriting coverage, sampling, non-response, and measurement errors present in these infrastructures. 

\eg{Persona Construction Error}{
\textbf{Definition. }The gap between the target population and the constructed sample of personas. 
\\
\\
\textbf{Example. }In~\citet{argyle2023out}, individual respondents from ANES serve as the simulation unit. The personas consist of several demographic and attitudinal attributes. Demographic characteristics have limited predictive power for many political attitudes and behaviors~\citep{kim2024division}. Finally, beyond the selection of persona variables, their expression might also impact simulation performance, e.g., saying ``You are a Democrat'' vs. ``You are liberal.''}

\subsection{Fielding of the Survey}

\textbf{LLM Selection and Inference. }For an LLM-based simulation, we now select the LLM to be used throughout the simulation, as well as its hyperparameters (model architecture, size, temperature, etc.). LLMs are complex systems shaped by both the data used to train them (e.g., how the pre-training and post-training data was sourced and curated), model architecture (e.g., decoder-based vs. encoder-decoder-based), and the different engineering steps in their creation (e.g., pre-training and post-training methods, data cleaning steps, harnesses used for Agentic LLM frameworks~\citep{yehudai2026survey}, etc).  We could also substitute regular LLMs with specialized versions, e.g., LLMs that have been fine-tuned on a specific population's data for better representativeness~\citep{ahnert2025extracting,suh2025language,cao2025specializing,kim2023ai,holtdirk2025learning} or LLMs specifically created for social research, e.g., Centaur, an LLM specialized to mimic human cognition~\citep{binz2024centaur}. Recent work has also specialized LLMs by manipulating their internal layers~\citep{jahanparast2026large}. LLM specialization need not always entail fine-tuning, steering via model internals, or customized models; it can also be done post-hoc via prompting, where examples or background knowledge is included in the prompt (``specialization'' in Figure~\ref{fig:prompt}(B)). Another design choice w.r.t. LLM selection entails choosing between instruction-tuned and base models; while the former type of models are easier to use for response generation, they may also be prone to reduced diversity~\citep{hu2025simbench,binz2026post}. 

The LLM can now be initialized with the prompt to simulate respondents. The LLM generates output based on this prompt, which can range from simple single-token responses for closed survey questions to complex free-text responses that describe the behavior of the entity they are simulating~\citep{ahnert-etal-2026-survey}. LLMs are imperfect simulators: their simulations of certain groups and certain constructs are worse, but not necessarily systematically~\citep{boelaert2025machine}. They also display response biases, e.g., recency bias~\citep{tjuatja2024llms}. Therefore, in LLM inference, we incur both measurement and representation errors, which emerge from the gaps between the \textit{actual} behavior of the real-world entity being simulated and the LLM-simulated behavior. To illustrate these errors due to imperfect LLM technology, we use the concept of an unbiased, error-free \textit{oracle LLM} that is free from `machine bias'~\citep{boelaert2025machine} and can perfectly simulate people; this oracle LLM acts as a proxy for human survey responses.\footnote{We assume human responses perfectly capture human behavior, though human surveys are themselves error-prone. For further discussion on this, see Section~\ref{sec:eval_errors}.} We differentiate this LLM-inherent error based on whether it impacts the measurement of the construct or the representation of a subgroup of the target population. 

\emph{\textbf{Response generation error}} is a measurement error that arises due to incorrect LLM responses to the task; it can have several causes: improper instruction-following,  response biases like social desirability, or guardrails~\citep{nudo2025generative} in LLMs. Response generation error is comparable to respondent-specific response errors (also called `measurement error' in~\citet{groves2005survey}) in the TSE, though past work shows that response biases of humans and LLMs do not fully align~\citep{tjuatja2024llms,boelaert2025machine,dominguez2024questioning,rupprecht_prompt_2025,dentella2023systematic}. Response errors in humans occur due to interactions between the survey question and human factors, including misinterpreting negated items, social desirability biases, or refusal to answer sensitive questions. Similarly, in LLM-simulated surveys, response generation errors occur due to the interaction between the question and LLM factors, including aspects of prompt design (c.f. Figure~\ref{fig:prompt}(B)) and the LLMs' inherent knowledge and synthesis gaps. 

Due to an LLM's stochastic nature, \textbf{reliability} is of great concern in LLM simulations, including survey simulations. LLMs might give different answers to the same prompt over multiple rounds, indicating low test-retest reliability. Test-retest reliability could be further compromised when using proprietary LLMs via Application Programming Interfaces (APIs) rather than a local open-source LLM; proprietary models might be updated, retrained, or even discontinued without announcement or documentation~\citep{barrie2024replication,mcloughlin2026proprietary}. Reliability is distinct from sensitivity~\citep{rothschild2026responsible}. Sensitivity, also used interchangeably with robustness~\citep{ye2026stopdrawingscientificclaims}, refers to changes in LLM output in the face of minute and inconsequential changes in input~\citep{sclar2023quantifying}, while reliability is measured by keeping the input identical.

\eg{Response Generation Error}{
\textbf{Definition. }Response generation error of a particular LLM is the gap between the ideal response to the task from an unbiased LLM and the obtained response.
\\
\\
\textbf{Example. }The LLM's answers to the vote choice question might be impacted by the order of the answer options~\citep{tjuatja2024llms}. Social desirability might also impact response generation error; LLM responses might have a higher share of a popular candidate. Certain LLMs might consider the request to pick an election candidate politically charged and refuse to answer.}

The representation error counterpart of response generation error is \emph{\textbf{persona simulation error}}. This is a representation error because the LLM misrepresents the respondent being simulated, e.g., due to demographic biases or stereotypes embedded in the LLM~\citep{cheng2023marked,wang2025large}. It is important to differentiate persona simulation error from persona construction error. While both are representation errors, the latter occurs upstream in the input phase and has to do with deficiencies in how the persona is constructed by the simulation designer. In contrast, persona simulation error is due to \textit{inherent} representational limitations in the LLM ---  that LLMs are either not trained on enough behavioral examples of \textit{all} potential groups of people or trained on misrepresentations of them~\citep{wang2025large}. Additionally, training processes might exacerbate representational harms, e.g., training data curation that only includes certain linguistic dialects~\citep{gururangan2022whose} and instruction-tuning that flattens opinion diversity~\citep{hu2025simbench,binz2026post}. The LLM can therefore not represent the simulated respondent regardless of how well the simulation designer constructs their persona. Current research has called into question whether LLMs are consistently good at simulating certain groups over others, labeling their idiosyncratic simulation performance as a novel type of `machine bias'~\citep{boelaert2025machine}.

Persona simulation errors can also occur due to interactions between prompt design, LLM selection, and inherent LLM factors while being compounded by persona construction errors. For example, how the persona is expressed as a part of persona construction impacts how well LLMs are able to simulate them~\citep{lutz2025prompt,weeber2026personacuesdifferentresults,tonneau2026differentdemographiccuesyield}. Persona simulation error is unique to LLM-simulated surveys; non-response error is the closest but imperfect counterpart in human surveys. While refusals can certainly be considered as LLM non-response, persona simulation errors also manifest as distortions and misrepresentations in LLM responses rather than outright refusals.

\eg{Persona Simulation Error}{
\textbf{Definition. }The gap between the ideal behavior of the simulated units, as obtained from an ideal error-free LLM, and the actual behavior of the simulated units obtained from the selected LLM.
\\
\\
\textbf{Example. }LLM-respondents might not simulate all groups of people equally well, e.g., for simulating American voters, LLMs might give systematically wrong answers for voters registered as Independents~\citep{argyle2023out}.}

\subsection{After Fielding the Survey}

\textbf{LLM Response Processing. }Now we process the LLM's responses into a format that is conducive for substantive analysis. For some use cases, this can be trivial, e.g., closed-ended surveys, where an LLM has to select one option out of a limited set. A researcher would simply count the responses and exclude any response that does not correspond to a valid answer option.\footnote{Invalid answers, e.g., refusals, are analogous to item non-response of human survey respondents and would be subsumed under response generation error in our framework.} Other use cases, e.g., for more complex survey tasks where LLM respondents generate comments, need more elaborate analysis pipelines. LLMs can also be used to process LLM-generated answers using the so-called `LLM-as-a-judge' approach~\citep{mellon2024ais}. As it is with human surveys, the type of preprocessing is often interlinked to the type of response solicited through the survey. For example, for LLM-generated surveys, if the designer asks the LLM to respond in JSON format, the response preprocessing would entail JSON parsing. The approach used in this response processing stage can be error-prone and lead to \emph{\textbf{response processing error}}: a measurement error due to the gap between the actual LLM response and the processed LLM response. Previous work has shown that processing steps can impact the overall simulation performance of LLMs in surveys~\citep{ahnert-etal-2026-survey,cummins2026threat}. Furthermore, when using stochastic processing workflows, e.g., LLM judges or other types of stochastic NLP models, test-retest reliability is also a concern at this stage. Response processing error is analogous to processing errors in the TSE.

\eg{Response Processing Error}{
\textbf{Definition. }The gap between LLM-generated responses and the processed responses.
\\
\\
\textbf{Example. }{For simulating American voters with an unrestricted prompt which simply asks ``Who will you vote for?'', we might get a range of answers, e.g., ``Trump'', ``I would vote for Trump'', ``The Republican candidate'', etc. One approach for processing these answers would be to do exact text matches with the full names of candidates. However, this exact match might be too conservative and remove instances where the LLM responded, for example, with only the first name, last name, a nickname or epithet.}}

\textbf{LLM Respondent Reweighting. }In addition to processing individual responses, the simulation designer may also adjust the \emph{contribution} of each LLM respondent to the final estimates. Post-survey adjustment is a common step in survey research, where survey designers use \textit{weights} based on auxiliary information to mitigate representation errors, especially coverage and non-response error. Adjustment can also be used to account for unequal selection probabilities, by aligning the achieved sample with known population benchmarks.  In LLM-generated surveys, one could also use post-study adjustment to control for persona simulation error if it is found that the LLM is particularly bad at simulating a particular group. Post-study adjustment or rectification is therefore an interesting option when combining real and simulated samples~\citep{broska2025mixed,krsteski2026valid}, where real samples might be substituted and upweighted for groups with a high degree of persona simulation error. However, just as in human surveys, where the adjustment stage might itself lead to representation errors due to incorrect weighting strategies~\citep{danielmeier2011post,groves2005survey}, the same is true for LLM-powered surveys. This type of error, called \emph{\textbf{adjustment error}}, can occur due to the choice of reweighting approach as well as the use of incorrect variables for weight creation. 

\eg{Adjustment Error}{
\textbf{Definition. }The gap between the statistic obtained from LLM-generated responses and the adjusted LLM responses obtained using an adjustment approach.
\\
\\
\textbf{Example. }The simulation designer could reuse the survey weights available from ANES for reweighting the LLM respondents. However, the efficacy of these weights is contingent on humans and LLMs having the same type of representation errors, which may not be the case.}

The simulation is now at the stage where we have all the LLM-simulated responses based on which we calculate the final simulated estimate, e.g., the vote choice of American voters. 

\section{Fallacies in Evaluating LLM-Generated Surveys}\label{sec:eval_errors}

Representation and measurement errors can be quantified (and potentially mitigated) when LLM-generated survey responses are contrasted against human-generated surveys, with the latter serving as a ground truth. Using human-generated survey data is the predominant approach for evaluating LLM-generated surveys~\citep{sen2025missing,vonderheyde2026AILoopSystematic}. However, this evaluation paradigm introduces several logical fallacies that we outline below.

\textbf{Ground Truth Fallacy} is concerned with the gap between survey-based measures and real-world behavior. Evaluating simulated survey responses against real survey data assumes that the latter is the `gold standard'; however, as the TSE framework demonstrates, survey data can also be a flawed source of social reality.
\textbf{Data contamination} can further impact the utility of survey data as ground truth, manifesting in two ways --- first, the LLM's training data might contain the survey responses that it should simulate~\citep{barrie_emergent_2025}, and second, the apparent human survey data could have been generated with partial or full assistance from an LLM~\citep{rilla2026recognising}. Data contamination can inflate the performance of an LLM's generated survey responses and this inflated performance might not predict the LLM's ability to simulate truly unseen contexts. Training data contamination errors are hard to establish for closed models whose training data and cut-off dates are unclear~\citep{barrie2024replication}. Furthermore, LLMs with web search capabilities might simply ingest survey results and reproduce them with high accuracy. 

\textbf{Single Best Metric Fallacy }refers to the use of one metric, or evaluation measure, to establish the performance of the LLM simulation. There is no consensus on the ideal metric for measuring simulation quality, also because different types of simulation might have different outputs and goals. For example, for closed-form survey responses, quantitative metrics like F1 score, Cohen's $\kappa$, or tetrachoric correlation can help assess match with human survey data on an individual-level, while distributional metrics like Wasserstein distance are used for group-level comparison~\citep{suh2025language,santurkar2023whose}. For open-ended generation, assessing quality can be more complex; such simulations might use construct-specific metrics such as toxicity of generated content~\citep{tornberg2023simulating}, or a wide variety of quantitative metrics like lexical diversity, tone, semantic similarity, and length ~\citep{pagan2025computational,zhang2025generative}. Qualitative metrics are also used, e.g, a human's ability to differentiate between real and LLM-generated responses~\citep{argyle2023out, bouleimen2025collective}.
Choosing the right metric for validating LLM-powered simulations is critical, but highly dependent on the specific simulation type, context, and purpose~\citep{gordon2026richer}. 
Finally, even when simulation performance is high, that might not translate to faithful results in downstream substantive analysis~\citep{barrie_cerina_2026}. Descriptive statistics are a means of establishing analytical patterns between variables; if the purpose of a simulation is multivariate modeling, we should also compare downstream inference obtained from the human-generated data and simulated data in terms of effect sizes and direction. 


\textbf{Context Drift Fallacy. }One of the most widely used approaches for applying LLM simulations involves first, validating simulated survey responses against some reference data and then applying it to a novel context without any reference data. This approach is called the `validate-then-simulate' approach~\citep{hullman2026human}. If the LLM's responses are found to be comparable to a reference using some type of evaluation metric, the designer could then use the LLM to simulate a scenario without validating against real human data. However, this paradigm falls victim to the \textbf{context drift fallacy } which refers to the fallacy of generalizing the LLM simulation's efficacy from the validated simulation setup to a supposedly `similar' context, e.g., validating LLM simulations on 2024 elections data and using the simulation's performance as evidence of its efficacy for a future election. It is still not clear \textit{if} two simulation contexts can be considered similar and \textit{how} one can quantify this similarity. Temporal drift is an issue in the quantitative social sciences in general;~\citet{munger2023temporal} describes `temporal validity' as a specific type of external validity or generalizability, where the target setting is in the future and is especially relevant in rapidly evolving social contexts such as studying digital media. For LLM-powered survey simulations, temporal drift is also noteworthy when simulating panel surveys. To simulate a future wave, a researcher would rely on information from current and past waves in creating artificial samples~\citep{krsteski2026valid}. For relatively stable attitudinal variables, using previous waves' data might not pose problems, but during volatile conditions, e.g., crises, temporal drift is high. Thus, further research is required to establish which periods and constructs are particularly prone to temporal drift, e.g., sudden pandemics~\citep{kozlowski2024simulating}. Similar concerns apply to geographic, cultural, or other context changes.

Temporal drift, like LLM response reliability, is also concerning when using closed-source models whose weights are unavailable; these are LLMs typically accessed via commercial APIs, e.g., ChatGPT or Claude. The provenance of such LLMs is unclear because changes to models are not publicly documented~\citep{barrie2024replication,aiyappa2023can}; furthermore, proprietary LLMs may be deprecated and made inaccessible without sufficient notice. Such changes threaten the reliability, validity, and reproducibility of LLM-generated survey simulations.

\section{Empirical Case Study: Simulating Survey Responses to the 2024 American National Election Survey (ANES)}
\label{sec:case_study}

We exemplify the Total Simulated Survey Error (TS2E) Framework using an empirical case study where we compare LLM-simulated responses against real human-generated survey responses. We simulate vote choices of American voters in the 2024 presidential election. To evaluate the LLM simulations, we use a reference dataset, i.e., the American National Election Studies (ANES) pre-election survey~\citep{anes_2024_2025}. 

\textbf{Design Choices.} To isolate the contribution of individual design choices to simulation quality, we follow a \textbf{multiverse analysis}~\citep{steegen2016increasing} approach: we systematically vary one design dimension at a time while holding others constant. The design dimensions we vary correspond directly to the error types identified in the TS2E framework, allowing us to assess the empirical magnitude of each error source in a controlled setting. Since we do not have an error-free oracle LLM, we cannot directly quantify the errors described in Section~\ref{sec:ts2e}. Instead, we use the reference data as a proxy of the result of an error-free LLM's response generation and assess how different design choices and different LLMs approximate this response generation.\footnote{Note that even with the reference data, we cannot measure the impact of certain design choices perfectly, simply because we do not have the reference data in response to those design choices, e.g., all possible question phrasings that we test in the simulation.}

We make and assess the impact of the following design choices: 

\begin{enumerate}
    \item \textbf{Task Articulation: }To obtain the vote choice of simulated American voters, we ask LLMs two versions of the same survey question: \textit{``Will you vote in the 2024 U.S. presidential election and if so for whom?''} \textbf{(v1)} and: \textit{``Please tell me if you will vote and if so for whom?''} \textbf{(v2)}. This allows us to assess \textit{articulation error}.
    \item \textbf{Persona Construction. }We reuse the sample and sample weights from the real ANES dataset, obtaining the information of 5,071 survey respondents who had indicated in the pre-election interview of the 2024 ANES Time Series Study that they voted or intended to vote for one of the two major candidates (\textbf{Trump} or \textbf{Harris}) or were unlikely to vote (labeled as \textbf{non-voters}). We convert each respondent into a simulation unit and design personas for each. The personas differ by composition and format: 
    \begin{enumerate}
        \item \textbf{Composition. }We assess the impact of two types of persona attributes. The first type of personas includes \textbf{demographics combined with attitudinal variables (Demo+Att)}, following Study 2 in \citet{argyle2023out}, which included variables based on their predictability of vote choice. These variables are: \textit{age, race, sex, state, attention to politics, ideology, party, attend church,} and \textit{campaign interest} (see Appendix~\ref{app:variables} for further details). The second type of personas includes \textbf{only demographics (Demo):} \textit{age, race, sex,} and \textit{state}.
        \item \textbf{Format. }We also have two variants of the persona prompt format — either an \textbf{interview-style} prompt~\citep{lutz2025prompt} or a declarative \textbf{direct} ``I am...'' format, giving us four different types of personas.
    \end{enumerate}  
    This allows us to assess \textit{persona construction error}.
    \item \textbf{LLM Selection. }We use six open-weight instruction-tuned models spanning three model families and two sizes each (see Appendix~\ref{app:reproducibility}): LLaMA-3.1-8B, LLaMA-3.3-70B, OLMo-3-7B, OLMo-3.1-32B, Qwen3-VL-8B, and Qwen3-VL-30B-A3B, which have been shown to perform well in vote prediction~\citep{ahnert-etal-2026-survey}. This allows us to assess the LLM-inherent component of \textit{response generation error}. By disaggregating the performance for different subgroups, we can also assess the LLM-inherent \textit{persona simulation error}.
    \item \textbf{Response Generation and Processing. }We vary the output mode of the prompt in two ways to obtain LLM responses: either a \textbf{JSON} response, extracted with structured outputs~\citep{dong2024xgrammar}, or an open-ended \textbf{free-text} response. The response processing method for the open-ended text response is varied in two ways: either a \textbf{regex}-based extractor that searches for heuristic keywords (e.g., ``Trump'' or ``Harris'') or an \textbf{LLM-judge}~\citep{li2024llms}. For the latter, we use GLM-4.7-Flash~\citep{5team2025glm45agenticreasoningcoding}, to extract the vote choice from the free-text answers.\footnote{To avoid `self-preference' biases in LLM judges~\citep{panickssery2024llm}, we do not use the GLM model as a response simulator LLM.} Since processing choices are often dictated by generation choices, we cannot fully disentangle the impact of either. Therefore, we collapse this into a single \textbf{response handling} category. Thus, we establish the impact of designer-driven \textbf{response generation} and \textbf{processing errors} jointly.
    \item \textbf{Adjustment. }We use the ANES survey weights, specifically the pre-election survey weights, to reweight the responses when computing the evaluation metrics (described in the following section). Thus, we have two further variations --- adjusted and non-adjusted, where all respondents are weighted equally in the latter. This allows us to assess \textit{adjustment error} as the impact of the adjustment process.
\end{enumerate}

\begin{wraptable}{r}{10cm}
\small
\centering
\begin{tabular}{llll}
\toprule
\textbf{Design Choice}                                                       & \textbf{Variant}           & \textbf{F1 ($\uparrow$)} & \textbf{TVD ($\downarrow$)} \\ \midrule
\multirow{2}{*}{Task}                                                        & \textbf{v1}                & \textbf{0.606 ± 0.14}    & \textbf{0.212 ± 0.09}       \\
                                                                             & v2                         & 0.604 ± 0.14             & 0.213 ± 0.09                \\ \midrule
                                                                             \multirow{4}{*}{Persona}                                                     & Demo (direct)              & 0.474 ± 0.04             & 0.268 ± 0.06                \\
                                                                             & Demo (interview)           & 0.465 ± 0.05             & 0.303 ± 0.07                \\
                                                                             & \textbf{Demo+Att (direct)} & \textbf{0.75 ± 0.04}     & \textbf{0.126 ± 0.03}       \\
                                                                             & Demo+Att (interview)       & 0.731 ± 0.05             & 0.153 ± 0.04                \\ \midrule
\multirow{6}{*}{LLM}                                                         & Llama-70B                  & 0.626 ± 0.13             & 0.208 ± 0.06                \\
                                                                             & Llama-8B                   & 0.593 ± 0.12             & 0.197 ± 0.07                \\
                                                                             & \textcolor{red}{Olmo-32B}                   & 0.619 ± 0.17             & \textcolor{red}{0.171 ± 0.09}       \\
                                                                             & Olmo-7B                    & 0.549 ± 0.15             & 0.267 ± 0.13                \\
                                                                             & \textcolor{blue}{Qwen-30B}                   & \textcolor{blue}{0.629 ± 0.14}    & 0.206 ± 0.08                \\
                                                                             & Qwen-8B                    & 0.614 ± 0.13             & 0.226 ± 0.08                \\ \midrule
                                                                             \multirow{3}{*}{\begin{tabular}[c]{@{}l@{}}Response\\ Handling\end{tabular}} & \textcolor{red}{JSON + Parsing}             & \textcolor{red}{0.615 ± 0.15}    & 0.227 ± 0.1                 \\
                                                                             & \textcolor{blue}{Free-text + LLM judge}      & 0.609 ± 0.15             & \textcolor{blue}{0.202 ± 0.09}       \\
                                                                             & Free-text + Regex          & 0.59 ± 0.13              & 0.209 ± 0.08                \\ \midrule
\multirow{2}{*}{Adjustment}                                                  & Adjusted                   & 0.595 ± 0.14             & 0.216 ± 0.09                \\
                                                                             & \textbf{Non-adjusted}      & \textbf{0.615 ± 0.15}    & \textbf{0.209 ± 0.1}        \\ \midrule
Overall                                                                      & Overall                    & 0.605 ± 0.14             & 0.212 ± 0.09                \\ \bottomrule
\end{tabular}
\caption{\textbf{Performance of Different Design Choices Measured with Weighted F1 (F1) and Weigted Total Variation Distance (TVD) scores.} We report mean performance with standard deviation. The best variant of each design choice is \textbf{bolded} if it is best based on both F1 and TVD, \textcolor{blue}{blue} if it is best by F1, and \textcolor{red}{red} if it is best based on TVD. }
\label{tab:config_table}
\end{wraptable}

This setup gives us 2 task versions $\times$ 4 personas $\times$ 6 LLMs $\times$ 3 response handling approaches $\times$ 2 adjustment approaches, bringing the total to 288 configurations. Each of the 144 pre-adjustment configurations ($2 \times 4 \times 6 \times 3$) are run over 5 random seeds to test simulation reliability of LLMs.\footnote{As adjustment comes after LLM simulation and is a deterministic process, we do not assess its reliability.} In total, we obtain 3,651,120 simulated responses (5,071 respondents $\times$ 144 pre-adjustment configurations $\times$ 5 runs). All prompts can be found in the Appendix (Section~\ref{app:prompts}) and the code for reproducing the simulation can be found here: \url{https://github.com/dess-mannheim/ts2e_anes_simulation}.

\textbf{Evaluation.} The reference data for the survey responses to be predicted are combined from \textit{turnout} and \textit{vote choice} in the ANES 2024 dataset (c.f. Appendix Section~\ref{app:variables} for further details). For each combination of design choices, we compute two metrics, with the ANES responses as our reference data: (1) weighted F1 score averaged over the three answer options (Harris, Trump, and non-voter) at the individual level and (2) weighted total variation distance (TVD) at the aggregate level. Here `weighted' does not refer to the use of survey weights, but weighing the metrics by the weight of classes of the predicted variable, i.e., group sizes of voters of Harris and Trump, and non-voters. We do this to account for the imbalance in the data. Higher F1 scores are better, while the opposite is true for TVD. To assess the impact of the aforementioned design choices on simulation performance, we use an OLS regression model where the dependent variable is the evaluation metric (F1 or TVD), while the independent variables are the five design choices. The number of observations per regression model are the 288 design choices. In addition to modeling the impact of design choices on overall performance, we also assess LLM-inherent \textbf{persona simulation error} by assessing the performance for ideological subgroups, e.g., F1 for conservatives, F1 for liberals, etc, with all design dimensions entered as categorical predictors.\footnote{We select ideology as an illustrative example given the vote choice task; other persona subgroups could also be used.} For the reference categories in the regression models, we use the best-performing configuration for that step based on overall performance (Table~\ref{tab:config_table}), per metric.

\begin{figure}
    \centering
    \includegraphics[width=0.98\linewidth]{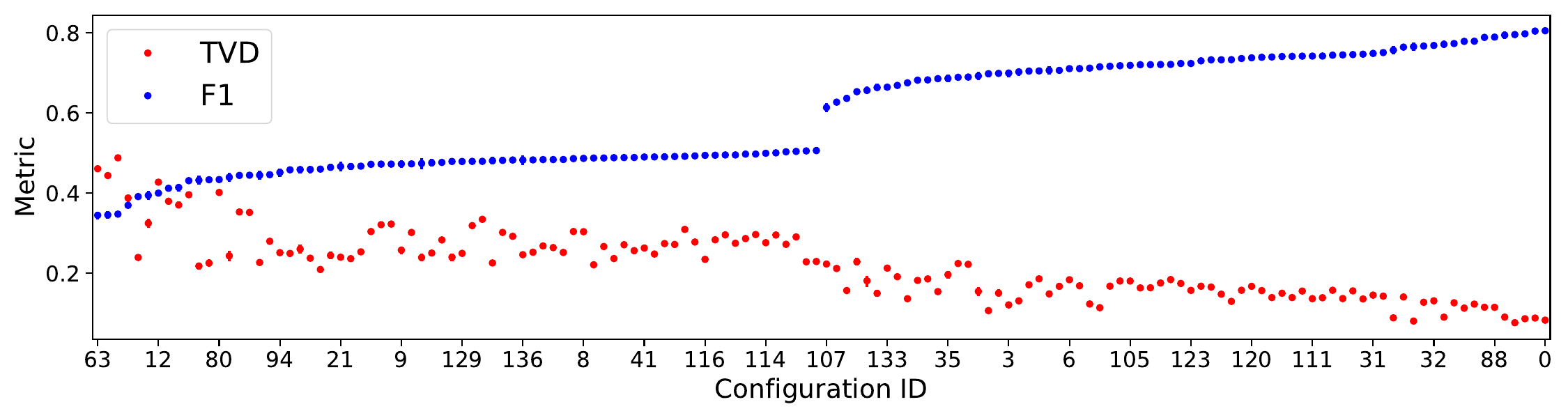}
    \caption{\textbf{Simulation Reliability of Design Choices.} We report mean scores for the two metrics (weighted F1 and TVD) of all 144 stochastic design configurations, sorted by F1 (only some configurations are labeled in the x-labels to ensure readability). The error bars show standard deviation of the metric. Almost all configurations are stable, and we generally find that configurations score similarly based on both metrics (lower TVD and higher F1 scores are better).}
    \label{fig:reliability}
\end{figure}

\textbf{Results.} Table~\ref{tab:config_table} shows the best-performing design choices as well as overall performance. Comparing the ranking of design configurations by F1 and TVD, we find high correlation ($\rho = -0.86, p < 0.001$).\footnote{The negative correlation is expected, since lower TVD scores indicate better performance.} However, the best configurations differ based on metric, e.g., the best response handling option based on F1 is JSON, while it is free-text and LLM-judge by TVD. For a given configuration, model responses are generally stable, indicating high reliability of all LLMs to the same input, i.e., reliability of 144 stochastic configurations over 5 seeds (Figure~\ref{fig:reliability}).

In addition to reliability, we also assess sensitivity, i.e., changes in results across different configurations. \textbf{We find a high degree of simulation sensitivity} --- there is a large difference between the best-performing configuration (F1 = 0.827, TVD = 0.060) and the worst-performing one (F1 = 0.338, TVD = 0.506). The best and worst configurations are included in the appendix (Table ~\ref{tab:best_worst_config} in Section~\ref{app:anes_results}). Conceptually, it might make sense that certain persona configurations are more informative, but the other design choices should ideally not play an outsized role for LLM performance. The fact that they do, points to the inherent brittleness of LLMs.

\begin{figure*}[h!]
    \centering
    \includegraphics[width=0.98\linewidth]{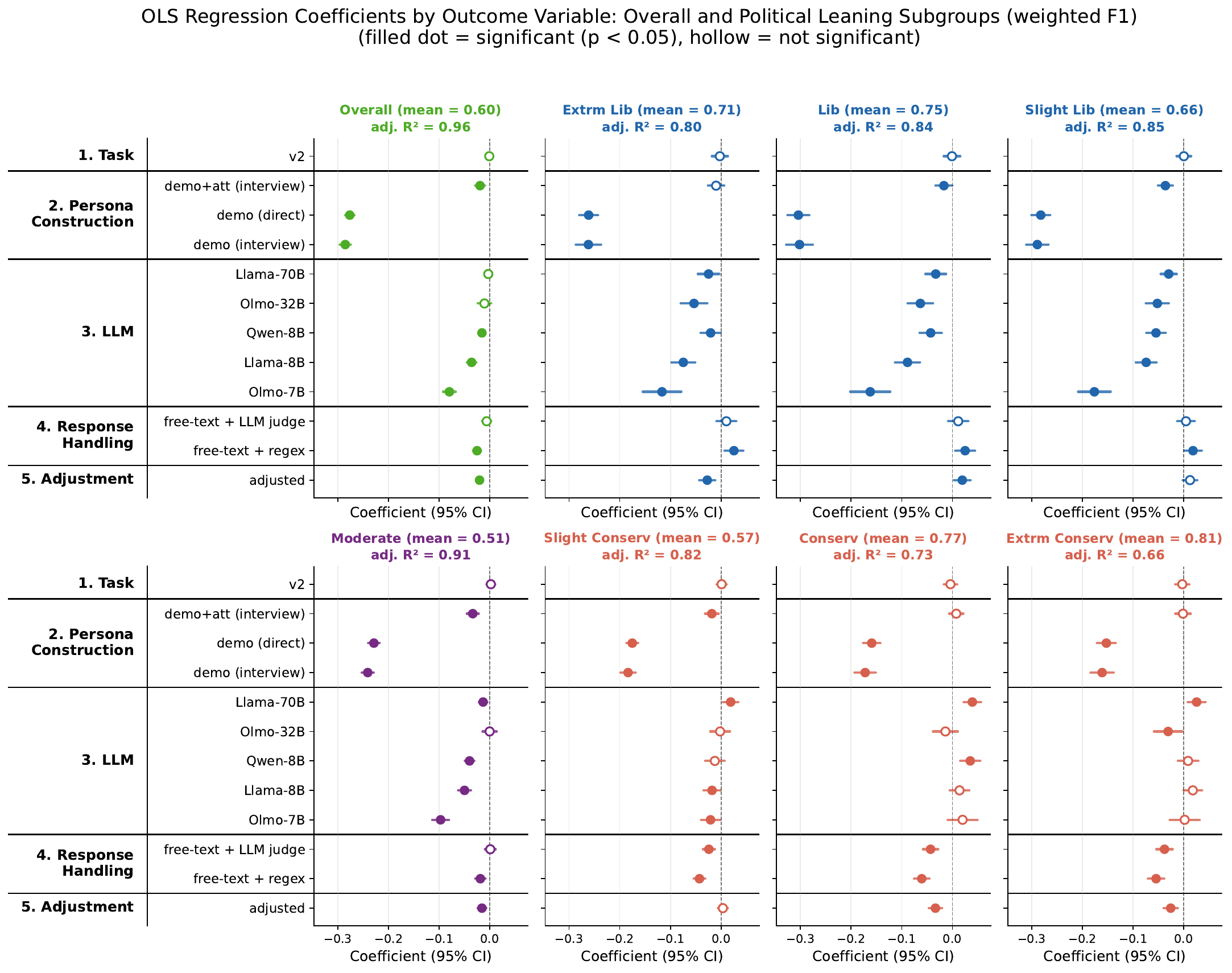}
    \caption{\textbf{Impact of simulation design choices on aggregate errors in simulating vote choice for the 2024 U.S elections.} OLS coefficients for different design choices in the simulation lifecycle (Y-axis) on weighted F1 score --- overall weighted F1 and F1 for the different ideological subgroups. 
    Reference categories for the regression are chosen based on the overall best-performing variant for a design choice. Therefore, all other design choices in the first subfigure (`overall') show that any listed design choice would \textit{reduce} F1 scores. However, when disaggregating by ideological subgroups, we find that for all conservative  subgroups (last three subfigures), certain LLMs (the LLaMa models and Qwen-8B) are better at predicting vote choice compared to the overall best-performing LLM (Qwen-30B). This differential performance for different subgroups surfaces \textbf{persona simulation errors} of LLMs.}
    \label{fig:regression_lollipop}
\end{figure*}

Figure~\ref{fig:regression_lollipop} reports the overall simulation performance and the simulation performance for different ideological groups (weighted F1, with the equivalent for weighted TVD in Figure~\ref{fig:regression_lollipop_tvd} of the Appendix Section~\ref{app:anes_results}). It also displays the OLS coefficients for each predictor of simulation performance, i.e., design choices, with rows ordered by their position in the design lifecycle of the TS2E. First, across all combinations, we see that the simulations achieve an overall weighted average F1 of 0.60, but the value changes based on the group, with higher performance for liberals and conservatives, but lower performance for moderates and slightly conservative subgroups. Second, we see that the adjusted $R^2$ is generally high, indicating that these design choices explain the variance in the simulation outcome. However, it is lower for moderates and conservatives. In appendix Section~\ref{app:anes_results}, we include a regression model with interactions between LLMs and design choice, which has high adjusted $R^2$ throughout (Figure~\ref{fig:interaction_F1} and~\ref{fig:interaction_TVD}). 

\textbf{Impact of Design Choices.} We do not find a statistically significant impact of task articulation on simulation performance. \textbf{The strongest and most consistent effect on overall performance as well as subgroup-disaggregated performance is the inclusion of attitudinal variables in the persona prompt, illustrating persona construction error.} Consistent with~\citet{argyle2023out}, adding certain attitudinal variables — which include party identification and ideology — produces a large gain, illustrating the importance of persona construction. 

Model choice also introduces model-specific biases in simulation quality. The LLM that performs best overall, i.e., Qwen-30B (the reference category in the regression), is not the best at predicting vote choice for conservative and extremely conservative ANES respondents, \textbf{thereby indicating that different LLMs might excel at simulating different subgroups.} Future \textit{in-silico} sampling set-ups could consider mixed LLM samples to leverage the differential simulation capabilities of different LLMs. Regarding response handling, we find that using free-text with regex parsing has a small but significant negative outcome on simulation performance. However, all the liberal groups fare better with this mode. Finally, surprisingly, \textbf{we find that using the ANES pre-survey weights for adjustments also has a small but significant negative impact}; only the liberal group benefits slightly from adjustment.

\textbf{Evaluation Fallacies. } 
We generally find a high correlation between the two metrics we use; however, they are not perfectly correlated. Notably, each metric ranks different LLMs as best. Ideally, the choice of metric should be justified by the use case for a simulation, while still reporting multiple appropriate metrics. Note that we measure alignment with the ANES, rather than with actual vote choice; in doing so, we run into the ground truth fallacy --- \textbf{ANES carries its own survey errors}. We avoid data contamination to some extent by using open-weight models whose cut-off dates are known and verified to be before the release of the ANES data. However, we cannot rule out that the survey data itself might be contaminated by respondents using LLMs.

We used the ANES 2024 survey data to construct personas and predict 2024 vote choice --- \textbf{both the persona variables and the target variable to be simulated are obtained from the same survey dataset.} However, if LLM-generated survey responses are to be useful in the real world, we would predict survey responses to a survey wave for which we do not already have the human survey responses for the target variable or where answers to the target variable are missing. In our case study, we find that personas with certain variables, including political ideology and party identification, improve prediction of the target variable, i.e., vote choice. It should be noted, however, that party identification and vote choice are strongly correlated. In the real world, answers to questions on both party preference and vote choice might be missing due to topical sensitivity; therefore, without the critical party identification variable, the prediction of vote choice by LLMs might be less accurate. Therefore, variables for persona construction would either need to be obtained from concurrent auxiliary sources, e.g., the census or past survey waves~\citep{krsteski2026valid}. Whether or not these variables are available or can inform vote choice in a different time setting invokes the context drift fallacy. For more realistic validations of LLM simulations, especially those that use fine-grained personas, \textbf{we need evaluation setups that mimic these real-world settings.} The setup would test an LLM's ability to predict survey responses in a context different to the one it is being instantiated in, e.g., a new time period or survey wave. These evaluation setups can be likened to out-of-domain evaluations in Machine Learning~\citep{wald2021calibration}.

\section{Related Work}

\subsection{Evaluation of Survey Data Quality in the Social Sciences}

\textbf{Survey Data Quality.} In survey research, evaluation of data quality focuses on whether the questionnaire measures the intended theoretical constructs in a valid and reliable manner across respondents, and whether the survey responses of the chosen sample generalize to an envisioned target population. Evaluations of surveys begin with the construct itself. Researchers ask whether the items actually capture the underlying concept, whether the wording is clear, and whether respondents interpret the questions in the intended way. 
On the other hand, survey quality is also evaluated based on representativeness of the survey sample, i.e., whether the target population within whom the construct is being measured has been adequately surveyed. Finally, the reliability of a survey is often assessed as the variable part of measurement error through internal consistency for multi-item scales, test-retest stability over time, and item-level diagnostics. 

\textbf{The Total Survey Error Framework.} The \emph{Total Survey Error} (TSE) framework provides a comprehensive approach to understanding, organizing, and evaluating all sources of error that arise in the production of survey statistics. Rather than focusing solely on a single type of error or the overall survey quality, TSE conceptualizes survey data quality as the cumulative result of multiple error sources that affect the quality of survey estimates~\citep{groves2010total}. The TSE framework classifies these error sources into \textbf{measurement errors} due to how the construct is defined, operationalized, and recorded, and \textbf{representation errors} due to how the target population is accessed and approximated. More details about the TSE framework can be found in Appendix Section~\ref{app:tse}.

A central strength of the TSE framework is that it offers ``a unified perspective on the design, conduct, and evaluation of surveys''~\cite[15]{groves2010total}. By decomposing error by survey stage, researchers can identify which processes contribute most to inaccuracy and prioritize interventions accordingly. 
The TSE framework also enables explicit trade-off analysis, recognizing that efforts to reduce one error source may inadvertently inflate another. A well-documented example is the non-response vs. response error trade-off. That is, intensive follow-up efforts to reduce non-response and refusal conversion increase response rates, but the reluctant respondents recruited in this way tend to provide lower-quality answers, for instance, through satisficing and, thereby, increasing response error~\citep{olson2006,fricker2010}. Survey design can, therefore, be approached as an optimization problem, minimizing total error across all components subject to cost constraints, rather than maximizing a single indicator such as the response rate~\citep{groves2010total}. Overall, the TSE framework provides a unifying conceptual framework for survey quality that supports both retrospective evaluation and prospective design by explicitly linking survey operations to statistical error components. For error sources under the researcher's control, it enables the diagnosis and reduction of specific errors linked to specific design choices; for error sources that cannot be fully controlled, it provides the means for quantifying, documenting, and (potentially) adjusting for their likely impact on survey estimates.

\subsection{Evaluation of LLM-powered Surveys}

A growing body of work evaluates how faithfully LLMs reproduce human attitudes and behavior, and a parallel body of work catalogs the ways in which they fail~\citep{peng2025mega,vonderheyde2026AILoopSystematic,desai2026validating}. On the evaluation side, efforts have moved from bespoke, study-specific comparisons toward standardized benchmarking: \citet{hu2025simbench, castricato-etal-2025-persona}, for instance, harmonize twenty datasets, including survey-based ones, into a single benchmark and find that even the strongest current models achieve only modest simulation fidelity, with performance degrading sharply when simulating specific demographic groups. Others have sought to systematize the types of validation rather than its scale; \citet{hullman2026human} distinguish several validation strategies---treating human and simulated subjects as interchangeable given prior evidence, directly comparing human and simulated response distributions, and Turing-style discrimination tests---and argue that each leads to different and limited inferential claims. 

As part of these evaluations, a critical literature also documents recurring failure modes of LLM-based simulations: social and cognitive biases, sycophancy, and ``alienness'' \citep{anthis2025llm}, as well as compressed response variance that distorts downstream estimates of uncertainty and effect size \citep{bisbee2024synthetic, doi:10.1177/08944393251337014, holtdirk_in-context_2026}, \textit{inter alia}.~\citet{desai2026validating} categorize different types of validation followed in LLM-based social science measurements, including LLM-generated survey responses, and find that validation practices are inconsistent and primarily focus on establishing convergent validity by comparing against human responses. 
\citet{larooij2025large} reach a similar conclusion in the adjacent field of LLM-powered (generative) agent-based models. They find that validation frequently rests on subjective assessments of ``believability,'' such that the opacity of LLMs may \emph{exacerbate} rather than resolve the long-standing validation challenges of simulation-based social science.

This literature establishes that LLM-simulated surveys are error-prone and offers increasingly rigorous ways to quantify their \textbf{overall error}, but it says comparatively little about \emph{why} these errors arise or \emph{where} in the simulation pipeline they originate. Guidelines from the American Association of Public Opinion Research (AAPOR) on the role of LLMs in the survey lifecycle discuss that validation of (human-generated) surveys goes beyond checking the validity of individual responses, but also tries to ensure the \textbf{traceability} and \textbf{transparency} of the entire survey lifecycle~\citep{rothschild2026responsible}; a similar expectation applies to LLM-generated survey data. Tools like the TSE framework are helpful in reflecting on the lifecycle of surveys. However, the TSE was not built for LLM-generated respondents. Further complications in process traceability arise due to difficulties in attributing an LLM's output to its training data and inference process. And even before honing into LLM-specific drawbacks, we lack an overview of where failures occur in the LLM-generated survey lifecycle. In particular, existing work rarely separates errors that stem from inherent limitations of current LLMs from those introduced by a researcher's design choices. Furthermore, we have yet to distinguish between errors that compromise \emph{what} is being measured from those that compromise \emph{whose} behavior is being represented \citep{heyde2025who}. This distinction between measurement and representation errors is consequential because the two call for different remedies. Diagnosing the origin of error, rather than merely measuring its overall magnitude, requires a framework that maps errors onto the specific steps of the simulation lifecycle at which they are introduced. We develop exactly such a framework, differentiating between measurement and representation errors and tracing each to its origin in the design lifecycle of an LLM-powered survey simulation.

\section{Discussion}

To obtain realistic survey simulations that closely mimic human behavior, a simulation designer's goal would be to minimize errors and reflect on the trade-offs between them. To that end, the Total Simulated Survey Error (TS2E) Framework describes representation and measurement errors at every stage of an LLM-powered survey simulation, some of which are consequences of the simulation designer's choices, others stemming from drawbacks in current LLM technology, and from the interplay between these.  We also discuss logical fallacies unique to LLM-powered survey simulations, where the evaluation of synthetic responses against human survey responses might pose further challenges. 

\textbf{Using TS2E.} Our framework can guide simulation designers in reflecting on the design choices throughout the simulation lifecycle, ideally even before starting a simulation as a \textbf{preregistration guide}~\citep{lakens2024benefits}. Pre-registering for the next LLM or survey wave has the potential to mitigate p-hacking risks~\citep{thomas2026mitigating} associated with the variance of model performance that we observe in our empirical case study in Section~\ref{sec:case_study}. Our framework can also help in the \textbf{design of benchmarks} for social simulations that focus on quantifying more nuanced errors rather than reporting a single performance score. Additionally, by focusing on errors at different stages of the simulation lifecycle, our framework can also inform \textbf{targeted interventions}, e.g., designing LLMs that are robust to differences in persona format. For the LLM-specific errors -- response generation and persona simulation -- tracing exactly where and why these errors emerge would require causal interpretability methods of LLMs and their training data. Such techniques are still being developed, while the training data of LLMs and the decisions in its curation are rarely publicly documented even for many open-source LLMs. Another complicating factor is that most \textit{in-silico} samples repurpose assistive instruction-tuned LLMs as simulators.\footnote{Some research does use base LLMs for simulations~\citep{suh2025language,jahanparast2026large}, but these are exceptions.} LLMs might have been pre-trained on a diverse range of data from different groups of people, including fringe groups. However, current post-training approaches that are used to create assistive LLMs might reduce the LLMs' capabilities in simulating fringe groups~\citep{binz2024centaur,hu2025simbench}, e.g., by introducing guardrails that prevent the LLM from discussing sensitive topics. Using aligned models for \textit{in-silico} sampling leads to a potential trade-off between realism and assistiveness. Thus, designing LLMs specialized for social simulations is an open interdisciplinary engineering challenge,  requiring input from social scientists as much as LLM designers; there is also the  question of \textbf{dual use} --- how would such specialized models, which might have different types of safety guardrails compared to existing assistive LLMs, be deployed and made available to the research community without being misused?

\textbf{Recommendations for LLM-Generated Survey Simulations. }In light of the errors we catalog, we urge simulation designers to carefully document the simulation context, their decision choices, as well as characteristics of the LLMs they use. We also suggest careful reflection on the validation of the LLM-simulated surveys and the contexts to which it generalizes, specifically, which respondent groups and time periods. When simulating past survey responses, we suggest caution in using proprietary closed-source LLMs whose training data provenance cannot be fully guaranteed, thus preventing safeguards against data contamination. In addition to establishing validity or the absence of errors, we also recommend assessing two further desiderata in evaluations --- (1) \textbf{reliability} which stresses changes in LLM output to the exact same input and design configurations and (2) \textbf{sensitivity} to inconsequential changes, e.g., the question phrasing. Frameworks such as QSTN by~\cite{kreutner2026qstn}, which allow for systematically varying different aspects of a simulation, can facilitate such assessments. For documentation, preregistration, and reflection, we suggest using the checklist in Appendix Section~\ref{app:checklist}, which extends existing checklists such as GuideLLM~\citep{feuerriegel2026reporting} and the AAPOR reporting standards~\citep{rothschild2026responsible} to include issues specific to survey simulation.

\textbf{\textit{In-Silico} Simulations Beyond Surveys.} LLMs can be used as proxies of people in various social scientific contexts, with survey responses being one of the pertinent uses cases. LLMs have also been used as stand-ins of people for content analysis~\citep{orlikowski2025beyond,mehrotra2026multi}, behavioral experiments~\citep{aher2022using,horton2023large}, experiments or evaluations that mimic human-LLM interactions~\citep{ivey2026real}, as well as simulations of behavior in offline and online systems~\citep{park2022social,park2023generative}. The latter are often described as ``generative agent-based models (GABMs)''~\citep{vezhnevets2023generative,larooij2025large}, that extend traditional agent-based models with LLMs or other generative AI tools. Many of the pitfalls that impact LLM-generated survey responses also impact these other use cases; the overarching question  remains the same--- can LLMs mimic human behavior faithfully? Therefore, certain errors in our framework can be applied without adaptation, e.g., persona construction and persona simulation. However, new sources of errors can emerge due to the specific context of these other simulations. For instance, feedback loops are a major concern in GABMs, especially when there are several simulation rounds. Future work can build specialized error frameworks for these other use cases.

\textbf{Ethical Considerations: Privacy and Autonomy.} Even if we were to overcome the technical limitations of current LLMs and succeed in designing as well as deploying LLMs that could simulate human populations, several ethical and epistemological concerns regarding this task would remain~\citep{agnew2024illusion,olteanu2025ai,crockett2023should}. Recent work has put forth frameworks for categorizing AI simulations of people, laying out important stakeholders for these systems~\citep{mcilroy2022mimetic,olteanu2025ai}. One crucial set of stakeholders are the people being simulated, e.g., the survey respondents who are being substituted by LLMs for filling out or finishing surveys on their behalf. Initial findings seem to indicate mixed responses~\citep{prolific2026participants}. Consent in LLM-generated surveys might be less of a concern when LLMs are used to simulate broad demographic groups (`Republicans'), but the closer we get to individuals, the thornier the issues of informed consent and predictive privacy become. Another issue is democratic legitimacy~\citep{cerina2025democratic}, especially when considering that public opinion informs policymaking. This warrants a discussion around whether and when agency over LLM-powered survey respondents should lie with LLM developers, survey designers, or the individuals being simulated~\citep{prolific2026participants,olteanu2025ai}. A lack of perceived legitimacy could also increase reactance or reluctance in human respondents to participate in surveys necessary for building, calibrating, and validating the simulations. Finally, if LLM-powered survey responses do turn out to be valid and reliable for certain contexts, human respondents might be replaced by LLMs.\footnote{There are several services available that attempt to simulate humans for market research, e.g., \url{https://simile.ai/blog/the-simulation-company} and \url{https://synthetic-humans.ai/}} What, then, happens \textbf{if we no longer have human-generated survey data to validate LLM-generated responses?}

\textbf{Limitations. }Given the vast design space of both human and \textit{in silico} surveys, and in order to ensure conciseness, our TS2E framework prioritizes breadth over depth. Our formal definition of human and LLM-generated surveys simplifies some steps, particularly post-survey adjustment since this is not main focus of our framework. The errors in our framework are based on stages in the \textit{in-silico} sampling lifecycle, but can have multiple finegrained subdimensions. For example, persona construction error can arise due to the selection of simulation units, the persona composition, as well as the persona format. Similarly, persona simulation errors can be due to undercoverage or misrepresentations in LLM pre-training data, LLM post-training data, or post-training processes. We hope future extensions of our framework can zoom into specific design steps and assess specific errors in greater depth. We also caution against over-interpreting the findings from the empirical use cases; they serve to illustrate the TS2E framework and do not generalize to the entirety of LLM-powered survey simulations. Indeed, given the wide variety of attitudes and behaviors that can be modeled with LLM simulations, it is possible that design choices that turned out to be statistically insignificant in our case study are significant in others and vice versa. Finally, our framework focuses on studies where the response sample is fully LLM-generated. Recent work on mixed designs that combine human and LLM responses~\citep{broska2025mixed,krsteski2026valid}, might introduce new types of errors depending on when and whose responses are simulated and how they are combined with human responses.

\section{Acknowledgments and AI Usage Disclosure}
We thank Ruggero Lazzaroni for his helpful comments on the validation of LLM simulations. Leah von der Heyde was supported by the Federal Ministry of Research, Technology and Space (BMFTR) under grant agreement 16DKZ2019A (KODAQS) and co-funded by the European Union – NextGenerationEU. Jana Lasser was supported by the European Research Council (ERC) under the European Union’s Horizon Europe program (Grant agreement No. 101160928). The authors used Claude Code to assist with designing some of the figures and robustness checks for the empirical case study. Claude Sonnet 5.5 and Writefull on Overleaf were used for proof-reading, grammar, and spelling checks. All authors read the manuscript and assume full responsibility for its content. 

\bibliography{references.bib} 

\section{Appendix}

This section comprises supplementary material for the paper ``Total Simulated Survey Error: Designing and Diagnosing Survey Responses from Large Language Models.'' We include further background on the Total Survey Error Framework that deals with the evaluation of human-generated survey responses (Section~\ref{app:tse}) and provide additional materials on the empirical case study in Section~\ref{sec:case_study} in the paper (Section~\ref{app:anes}) and the checklist for documenting simulated survey responses (Section~\ref{app:checklist}). 

\subsection{Description of the Total Survey Error Framework}\label{app:tse}

\begin{figure}[t!]
\begin{center}
\includegraphics[scale=0.5]{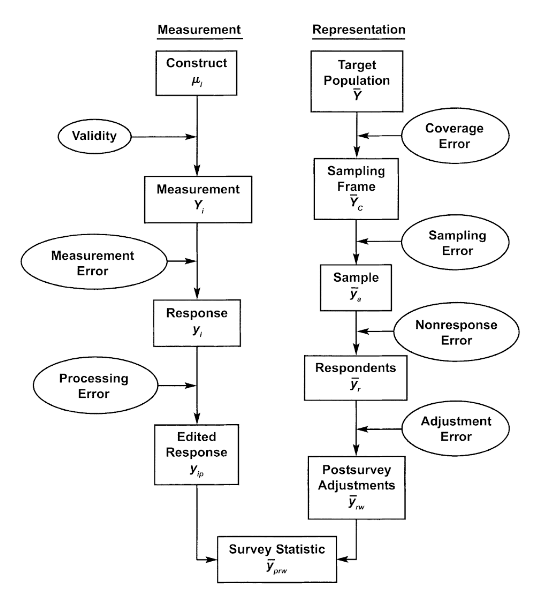}
\end{center}
\caption{\textbf{Total Survey Error Components} linked to steps in the measurement and representational inference process~\citep{groves2005survey}}
\label{fig:tse}
\end{figure}

The Total Survey Error (TSE) framework provides a unified view on assessing random and systematic errors in human-generated survey data~\citep{groves2005survey,biemer2010total}. Its diagrammatic representation can be found in Figure~\ref{fig:tse} 

In survey research, errors can be described by assessing the \textit{source} of errors, most prominently the distinction between \textbf{measurement} errors and \textbf{representation} errors \citep{groves2005survey}. The term `error' refers to the difference between the obtained value and the true value we want to measure, while bias refers to systematic errors, following the definitions in Weisberg's `The Total Survey Error Approach', p.22~\citep{weisberg2009total}. Measurement errors relate to defining and measuring a theoretical construct with chosen indicators \textit{(measurement errors}). Representation errors, on the other hand, are the errors arising when inferring from a sample to the target population. This distinction is helpful in conceptually disentangling different fallacies plaguing surveys as well as related research designs using social data. To that end, the TSE lens has been extended to other forms of social data~\citep{amaya2020total,boeschoten2020digital,sen2021total,bosch2022survey,schneck2025meta}.

The first step of a survey requires defining the theoretical construct of interest and establishing a theoretical link between the questionnaire that will measure this theoretical construct and the construct, itself~\citep{Howison2011}. Survey researchers usually start by defining the main construct of interest (e.g., ``sexist attitude'') and potentially related constructs (e.g., ``gender bias''). They then design or reuse survey scales (i.e., sets of questions and items) that will be used to \textit{measure the construct adequately}, establishing \textit{validity}. In developing scales, content validity, convergent construct validity, discriminant construct validity, internal consistency, as well as other quality marks are checked (cf.~\citet{straub2004validation}). \citet{groves2005survey} further point out ``measurement error'' (not to be confused with the measurement error
arm of the TSE), which is also called \textit{response error}~\citep{hansen1951response}. Response errors arise during the solicitation of actual information in the field, even when an ideal measurement has been found. A respondent may not understand a set of items as intended, may not be able (usually through recall problems) or may not want to (e.g., social desirability) answer the question truthfully or at all. Finally, \textit{processing errors} can be introduced when processing data, such as coding textual answers into quantitative indicators or data cleaning. In addition to validity, individual responses may also suffer from variability over time or between participants, contributing to low \textit{reliability}. 


Survey attempt to obtain an unbiased representation  of their target population by clearly defining the it; this is the population in which the construct of interest should be measured for/inferred to, e.g., the national population of a nation-state. Then, a sampling frame is defined, i.e., the best approximation of all units in the target population, e.g., telephone lists or (imperfect) population registers. By choosing such a sampling frame, under- or over-coverage of population elements might occur, e.g., some people being left out because their phone numbers are not listed in a telephone list or they do not have a phone, and some people being counted twice because they have multiple phone numbers. Therefore, drawbacks of the sampling frame lead to \textit{coverage error}. Mostly due to financial constraints, but also due to logistical infeasibility, it is not possible to survey every element in the sampling frame. By ignoring these elements, the sample statistics will most likely deviate from the (unobservable) sampling frame statistics, thereby introducing a \emph{sampling error}. The sampling error can be decomposed into two components: sampling variance, the random part, and sampling bias, the systematic part. The sampling variance is a consequence of randomly drawing a set of elements from a sampling frame. Sampling bias as the second component of sampling error comes into play when the sampling process is designed and/or executed in such a way that a subset of units is selected from the sampling frame but giving some members of the frame systematically lower chances to be chosen than others. 

Further, if chosen individuals drawn as part of the sample refuse to answer the whole survey, we speak of unit \textit{non-response errors}. While in most cases, providing insufficient responses to items hinders valid inferences regarding the topical research questions, non-response to demographic items can also hinder post-survey adjustment of representation errors.\footnote{While not mentioned explicitly by~\citet{groves2005survey}, this affects the adjustment step and becomes much more important when working with digital traces.} Lastly,~\citet{groves2005survey} list \textit{adjustment error}, occurring when reweighting is applied post-survey to under- or over-represented cases due to any of the representation errors described above. The reweighting is usually based on socio-demographic attributes of individuals and often their belonging to a certain stratum. Importantly, adjustment steps do not necessarily fully fix representation errors; indeed, they can also introduce new ones.

\subsection{Empirical Simulation: 2024 ANES Simulation: Experimental Setup}\label{app:anes}

\subsubsection{Variable Selection} \label{app:variables}

All 2024 ANES variables that we used in the simulation study are documented in Table~\ref{tab:anes_variables}. We stayed as close as possible to the variables selected in Study~2 from \citet{argyle2023out}. Race/ethnicity, age, ideology, party identification, church attendance, and state are direct 2024 counterparts of their items, and sex (\texttt{V241550}) corresponds to the item they used for 2020. As in \citet{argyle2023out}, if a persona variable is missing for a respondent, the corresponding template fragment is omitted from the persona. Our variables differ from theirs in four ways, \textbf{most critically, to update the setting to 2024 and to reduce mixing pre- and post-election variables}. First, because we use pre-election variables only, we replace their two post-election items: interest in politics becomes interest in the political campaigns (\texttt{V241005}), and discussing politics with family and friends becomes attention to politics and elections (\texttt{V241004}), since the pre-election interview does not ask respondents about their own political discussions. Second, we omit their 2016 item on feelings about the American flag, which is not part of the 2024 ANES and was also not asked in 2020. Third, they predict the self-reported vote as a binary choice between the Democratic and Republican candidates, whereas we predict pre-election vote intention and add Non-voter as a third category. Fourth, we use survey weights in the weighted evaluation metrics, whereas \citet{argyle2023out} report unweighted metrics only.

Our simulation target, vote choice (Harris, Trump, Non-voter), combines \textbf{vote intention} and \textbf{turnout likelihood}. Respondents who had already voted early (\texttt{V241075x} = 10/11) keep the candidate they voted for, regardless of their turnout likelihood. All others who are only slightly or not at all likely to vote (\texttt{V241102x} = 5/6) are coded as Non-voters, the remaining respondents by their intended vote (\texttt{V241075x} = 20/21). The paper questionnaire has no vote intention question, so for paper respondents (\texttt{V240002a} = 3) the candidate preference (\texttt{V241075x} = 30/31) is used instead. The remaining respondents who voted for or intend to vote for a candidate other than Trump or Harris, who only state a preference in non-paper interview modes, or who have missing values are excluded (final sample: $N=5,071$). The weighted evaluation metrics use the pre-election weight \texttt{V240107a}.

\begin{table}[ht]
  \centering
  \setlength{\abovecaptionskip}{5pt}
  \begin{threeparttable}
    \begin{tabular}{@{}lp{.46\textwidth}lrcc@{}}
    \toprule
    Variable & Description & Usage & \% Missing\tnote{a} & Demo+Att & Demo \\
    \midrule
    \texttt{V241458x} & Age on election day & Persona & 5.1 & \checkmark & \checkmark \\
    \texttt{V241501x} & Self-identified race/ethnicity & Persona & 1.2 & \checkmark & \checkmark \\
    \texttt{V241550} & Sex & Persona & 0.8 & \checkmark & \checkmark \\
    \texttt{V243001} & Sample location state & Persona & 0.0 & \checkmark & \checkmark \\
    \texttt{V241004} & Attention to politics and elections & Persona & $<$0.1 & \checkmark & \\
    \texttt{V241177} & 7-point liberal-conservative self-placement & Persona & 14.3\tnote{b} & \checkmark & \\
    \texttt{V241227x} & Party ID (7-point summary) & Persona & 0.7 & \checkmark & \\
    \texttt{V241439} & Ever attends church or religious services & Persona & 5.1 & \checkmark & \\
    \texttt{V241005} & Interest in following the political campaigns & Persona & 4.5 & \checkmark & \\
    \midrule
    \texttt{V241102x} & How likely the respondent is to vote & Outcome & $<$0.1 & -- & -- \\
    \texttt{V241075x} & Party of pre-election presidential vote/intent/preference & Outcome & 7.2\tnote{c} & -- & -- \\
    \midrule
     & Vote choice (aggregated) & Outcome & 8.2\tnote{d} & -- & -- \\
    \midrule
    \texttt{V240107a} & Pre-election full sample weight & Weight & 0.0 & -- & -- \\
    \bottomrule
    \end{tabular}
    \begin{tablenotes}\footnotesize
      \item[a] Share of all respondents who completed the pre-election interview ($N=5,521$).
      \item[b] The answer ``haven't thought much about this'' is labeled as missing.
      \item[c] Respondents who are slightly or not at all likely to vote (5.1\%) are coded as Non-voters.
      \item[d] Excluded respondents: 5.8\% support a candidate other than Harris or Trump, and 2.4\% give no codable vote choice.
    \end{tablenotes}
    \caption{Variables from the 2024 ANES Time Series Study used in the case study. All variables come from the pre-election interview. Persona variables are used to construct the LLM personas, and check marks indicate which variables each persona type (\textbf{Demo+Att} or \textbf{Demo}) includes. Missing values of persona variables are omitted from the persona prompt, i.e., the sentence expressing that field is excluded. The outcome variables are combined into the vote choice target (Harris, Trump, Non-voter). The sample weight is used for the adjusted evaluation metrics.}
  \end{threeparttable}
  \label{tab:anes_variables}
\end{table}

\subsubsection{LLM Hyperparameters and Reproducibility} \label{app:reproducibility}

\textbf{Hardware. }All simulations and annotations were run on a cluster with two GPU nodes: one with NVIDIA RTX PRO 6000 Blackwell Max-Q GPUs and one with NVIDIA H100 PCIe GPUs. Each job used a single GPU, except for Llama-70B, which used two GPUs on the same node. Jobs were assigned to nodes by availability, so the random seeds of the six models ran on a mix of both GPU types. Each simulation run (one model and seed) took between 20 minutes and 2.4 hours, and the full set of simulations and annotations took 45.1 GPU-hours, or 10.1 hours of wall-clock time.

\textbf{Software. }We ran all simulations and annotations with the QSTN framework~\citep{kreutner2026qstn} (v0.3.3), using vLLM (v0.21.0) as the backend for local model inference. The huggingface model ids and model revisions are shown in Table~\ref{tab:HF_models}. The code for our simulation, annotation, and evaluation pipeline, along with the simulated responses and aggregated results, is available at \url{https://github.com/dess-mannheim/ts2e_anes_simulation}. The ANES data is not included and can be obtained from \url{https://electionstudies.org/data-center/2024-time-series-study/}.

\begin{table}[t]
    \centering
    \footnotesize
    \setlength{\abovecaptionskip}{10pt}
    \begin{tabular}{lll}
    \toprule
    \textbf{Shorthand} & \textbf{Model} & \textbf{Hugging Face revision} \\
    \midrule
    LLaMA-8B & \href{https://huggingface.co/meta-llama/Llama-3.1-8B-Instruct}{\texttt{meta-llama/Llama-3.1-8B-Instruct}} & \texttt{0e9e39f249a16976918f6564b8830bc894c89659} \\
    LLaMA-70B & \href{https://huggingface.co/meta-llama/Llama-3.3-70B-Instruct}{\texttt{meta-llama/Llama-3.3-70B-Instruct}} & \texttt{6f6073b423013f6a7d4d9f39144961bfbfbc386b} \\
    OLMo-7B & \href{https://huggingface.co/allenai/Olmo-3-7B-Instruct}{\texttt{allenai/Olmo-3-7B-Instruct}} & \texttt{6e5971d9eba42665f5bd5a0fcf047f299ce1dccc} \\
    OLMo-32B & \href{https://huggingface.co/allenai/Olmo-3.1-32B-Instruct}{\texttt{allenai/Olmo-3.1-32B-Instruct}} & \texttt{ac0587e4a7744a551c059d8cd17ba220bc940dae} \\
    Qwen-8B & \href{https://huggingface.co/Qwen/Qwen3-VL-8B-Instruct}{\texttt{Qwen/Qwen3-VL-8B-Instruct}} & \texttt{0c351dd01ed87e9c1b53cbc748cba10e6187ff3b} \\
    Qwen-30B & \href{https://huggingface.co/Qwen/Qwen3-VL-30B-A3B-Instruct}{\texttt{Qwen/Qwen3-VL-30B-A3B-Instruct}} & \texttt{9c4b90e1e4ba969fd3b5378b57d966d725f1b86c} \\
    GLM-4.7-Flash & \href{https://huggingface.co/zai-org/GLM-4.7-Flash}{\texttt{zai-org/GLM-4.7-Flash}} (annotation) & \texttt{7dd20894a642a0aa287e9827cb1a1f7f91386b67} \\
    \bottomrule
    \end{tabular}
    \caption{Hugging Face models and the revisions used.}
    \label{tab:HF_models}
\end{table}

\textbf{Hyperparameters. }All models used vLLM's default sampling parameters (temperature 1.0, top-p 1.0, no top-k, repetition penalty 1.0). We limited outputs to 7,000 tokens and the context length to 15,000 tokens, disabled thinking for GLM-4.7-Flash via the chat template, and gave each request its own seed derived from the run seed.

\newpage
\subsubsection{Simulation Prompts}\label{app:prompts}

We list all the exact prompts used in the empirical case study.

\promptmsg{promptsysborder}{System prompt for \textit{free-text} response generation}{
You are a political scientist predicting vote choice in the 2024 U.S. presidential election.
}

\promptmsg{promptsysborder}{System prompt for \textit{JSON} response generation}{
You are a political scientist predicting vote choice in the 2024 U.S. presidential election. You only respond with the most probable answer option in the following JSON format:\\
\{\\
\ \ "answer": "choose one of: Options are: Harris, Trump, Non-voter"\\
\}
}


\promptmsg{promptuserborder}{User prompt for \textit{first-person} personas and \textit{Task v1} (example)}{
Racially, I am \promptvar{Black}. I \promptvar{always} pay attention to what's going on in government and politics. Ideologically, I am \promptvar{moderate}. Politically, I am \promptvar{an independent who leans Democratic}. I \promptvar{do not} attend church. I am \promptvar{29} years old. I am \promptvar{female}. I have been \promptvar{somewhat} interested in the political campaigns so far this year. I am from \promptvar{South Carolina}. When I'm being asked `\promptvar{Will you vote in the 2024 U.S. presidential election and if so for whom?} Options are: Harris, Trump, Non-voter', I respond with
}

\promptmsg{promptuserborder}{User prompt for \textit{first-person} personas and \textit{Task v2} (example)}{
Racially, I am \promptvar{Black}. I \promptvar{always} pay attention to what's going on in government and politics. Ideologically, I am \promptvar{moderate}. Politically, I am \promptvar{an independent who leans Democratic}. I \promptvar{do not} attend church. I am \promptvar{29} years old. I am \promptvar{female}. I have been \promptvar{somewhat} interested in the political campaigns so far this year. I am from \promptvar{South Carolina}. When I'm being asked `\promptvar{Please tell me if you will vote and if so for whom.} Options are: Harris, Trump, Non-voter', I respond with
}

\promptmsg{promptuserborder}{User prompt for \textit{interview} personas and \textit{Task v1} (example)}{
Interviewer: What is your race?\\
Interviewee: I am \promptvar{Black}.\\
Interviewer: How often do you pay attention to what's going on in government and politics?\\
Interviewee: I \promptvar{always} pay attention to what's going on in government and politics.\\
Interviewer: What is your political ideology, from extremely liberal to extremely conservative?\\
Interviewee: Ideologically, I am \promptvar{moderate}.\\
Interviewer: Which party do you identify with politically?\\
Interviewee: Politically, I am \promptvar{an independent who leans Democratic}.\\
Interviewer: Do you ever attend church?\\
Interviewee: I \promptvar{do not} attend church.\\
Interviewer: How old are you?\\
Interviewee: I am \promptvar{29} years old.\\
Interviewer: What is your gender?\\
Interviewee: I am \promptvar{female}.\\
Interviewer: Would you say that you have been very much interested, somewhat interested or not much interested in the political campaigns so far this year?\\
Interviewee: I have been \promptvar{somewhat} interested in the political campaigns so far this year.\\
Interviewer: Which state are you from?\\
Interviewee: I am from \promptvar{South Carolina}.\\
Interviewer: \promptvar{Will you vote in the 2024 U.S. presidential election and if so for whom?} Options are: Harris, Trump, Non-voter\\
Interviewee:
}

\promptmsg{promptuserborder}{User prompt for \textit{interview} personas and \textit{Task v2} (example)}{
Interviewer: What is your race?\\
Interviewee: I am \promptvar{Black}.\\
Interviewer: How often do you pay attention to what's going on in government and politics?\\
Interviewee: I \promptvar{always} pay attention to what's going on in government and politics.\\
Interviewer: What is your political ideology, from extremely liberal to extremely conservative?\\
Interviewee: Ideologically, I am \promptvar{moderate}.\\
Interviewer: Which party do you identify with politically?\\
Interviewee: Politically, I am \promptvar{an independent who leans Democratic}.\\
Interviewer: Do you ever attend church?\\
Interviewee: I \promptvar{do not} attend church.\\
Interviewer: How old are you?\\
Interviewee: I am \promptvar{29} years old.\\
Interviewer: What is your gender?\\
Interviewee: I am \promptvar{female}.\\
Interviewer: Would you say that you have been very much interested, somewhat interested or not much interested in the political campaigns so far this year?\\
Interviewee: I have been \promptvar{somewhat} interested in the political campaigns so far this year.\\
Interviewer: Which state are you from?\\
Interviewee: I am from \promptvar{South Carolina}.\\
Interviewer: \promptvar{Please tell me if you will vote and if so for whom.} Options are: Harris, Trump, Non-voter\\
Interviewee:
}

\subsubsection{LLM Judge Prompts}

\promptmsg{promptsysborder}{System prompt for the annotation of free-text responses with an LLM judge}{
You are an expert annotator labeling statements about vote choice in the 2024 U.S. presidential election. You only respond with the most probable answer option in the following JSON format:\\
\{\\
\ \ "answer": "choose one of: Options are: Harris, Trump, Non-voter"\\
\}
}

\promptmsg{promptuserborder}{User prompt for the annotation of free-text responses with an LLM judge}{
Below is a STATEMENT that responds to the question: "Will you vote in the 2024 U.S. presidential election and if so for whom?"
Does the following STATEMENT express preference for exactly one of the following answer options? Harris, Trump, Non-voter. If yes, which option is preferred?\\
STATEMENT: \promptvar{\{LLM response\}}}

\newpage
\subsection{Empirical Simulation: Further Results}\label{app:anes_results}

\subsubsection{Best and Worst Configurations}

The best and worst-performing design configurations are listed in Table~\ref{tab:best_worst_config}.

\begin{table}[]
\centering
\footnotesize
\setlength{\abovecaptionskip}{10pt}
\begin{tabular}{llllllll}
\toprule
\textbf{Config ID} & \textbf{Response Handling} & \textbf{LLM} & \textbf{Task} & \textbf{Persona}     & \textbf{Adjustment} & \textbf{F1} & \textbf{TVD} \\
\midrule
237                & JSON + Parsing              & Olmo-32B     & v2            & Demo+Att (interview) & Non-adjusted        & \textbf{0.827}       & 0.070        \\
282                & Free-text + LLM judge       & Olmo-32B     & v2            & Demo+Att (interview) & Non-adjusted        & 0.820       & \textbf{0.060}       \\
256                & Free-text + LLM judge       & Olmo-7B      & v2            & Demo (interview)     & Non-adjusted        & \textit{0.338}       & 0.461        \\
146                & JSON + Parsing              & Olmo-7B      & v1            & Demo (direct)        & Non-adjusted        & 0.342       & \textit{0.506} \\
\bottomrule
\end{tabular}
\caption{\textbf{Best} and \textit{worst} configurations in the ANES simulation study based F1 scores and TVD.}
\label{tab:best_worst_config}
\end{table}

\subsubsection{Predicting Vote Choice of Different Ideological Groups}

\begin{figure}
    \centering
    \includegraphics[width=0.95\linewidth]{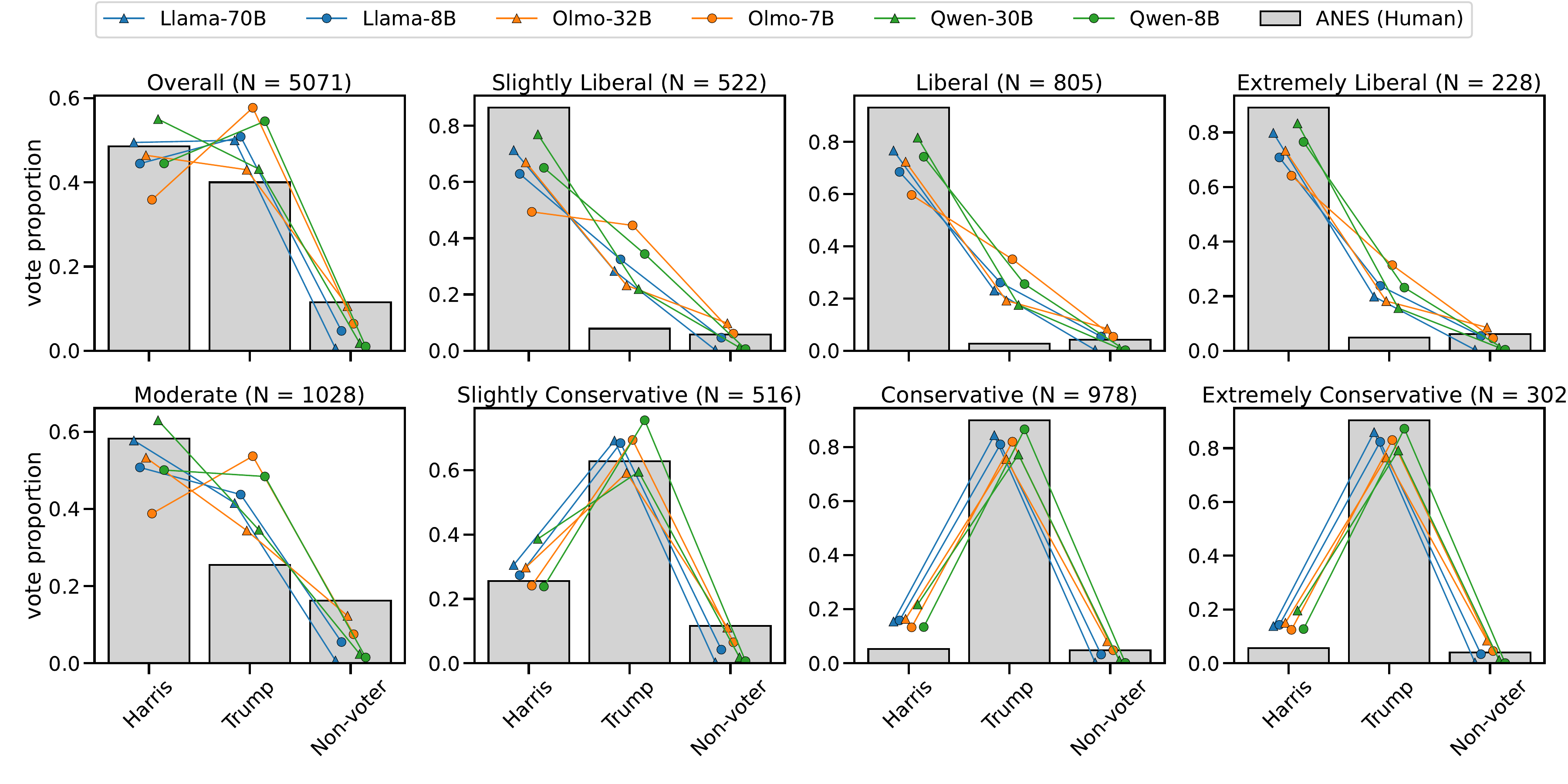}
    \caption{\textbf{Vote Share in ANES vs. LLM Predictions.} LLMs generally underpredict non-voters for all different ideologies.}
    \label{fig:persona_simulation_error}
\end{figure}

Figure~\ref{fig:persona_simulation_error} shows the performance of LLMs for the different ideology subgroups and helps contextualize persona simulation error and the regression findings in Figure~\ref{fig:regression_lollipop}. Specifically, we see that the model that performs best overall,  Qwen-30B w.r.t. to F1 scores, heavily underestimates non-voter rates whose share is relatively high in the moderates and slightly conservative subgroups. 

\subsubsection{Total Variation Distance and Interactions}

Complementing the results in Section~\ref{sec:case_study}, we include two additional analyses: (1) simulation results measured with Total Variation Distance (TVD) and (2) a regression model with interactions between different design choices. Finally, we also include all the prompts used to simulate voting behavior with LLMs (Section~\ref{app:prompts}).

\begin{figure*}
    \centering
    \includegraphics[width=0.98\linewidth]{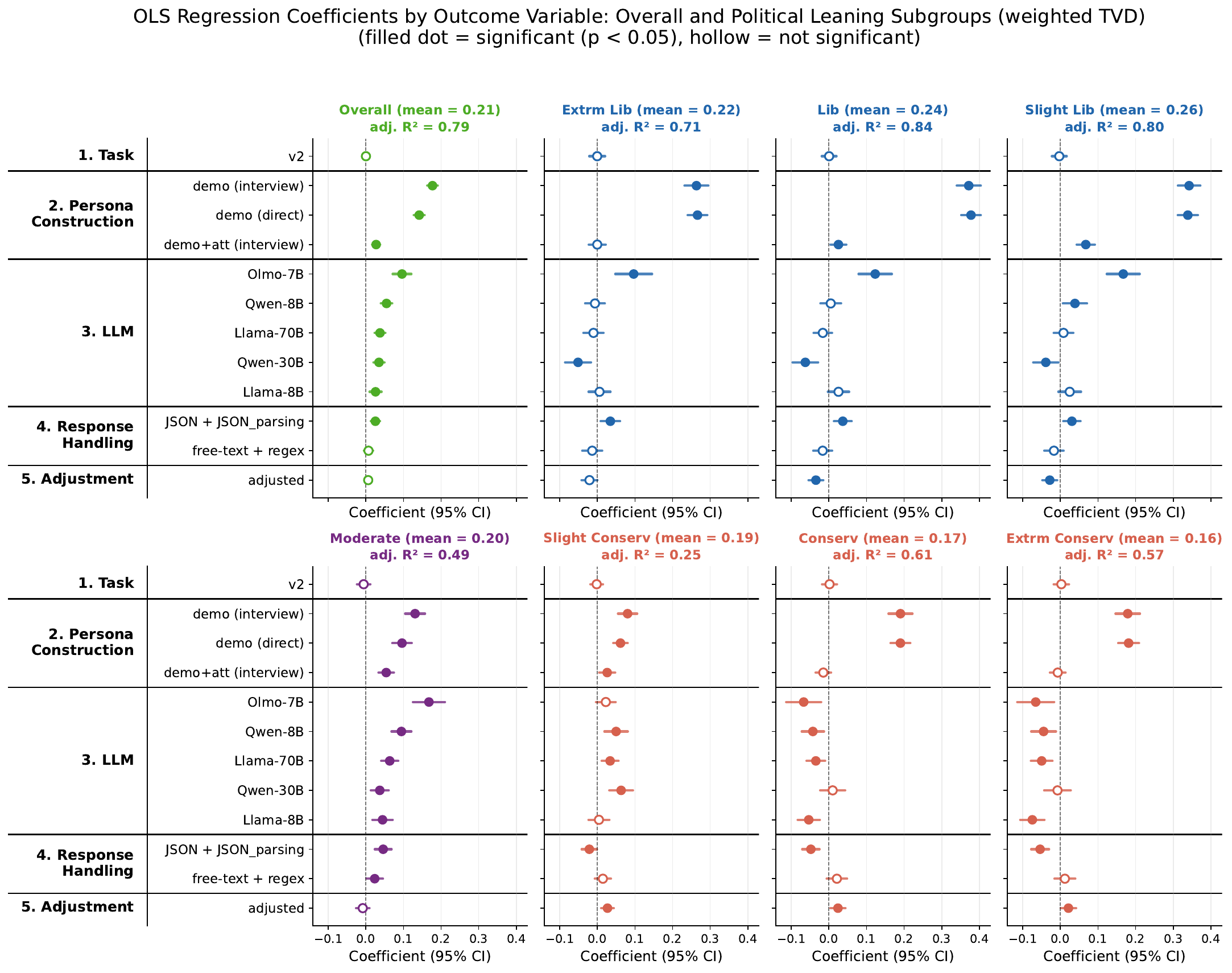}
    \caption{\textbf{Impact of simulation designer choice on aggregate errors in simulating vote choice for the 2024 U.S elections.} OLS coefficients for different design choices in the simulation lifecycle (Y-axis) on weighted TVD score --- overall weighted TVD and TVD for the different ideology subgroups. Error bars represent standard errors. Compared to F1 scores, we see a lower amount of variance explained by design choices for moderate and conservative groups. Since lower TVD scores are better, the reference variants lead to the best overall TVD and any other variant would \textit{increase} TVD. Similar to Figure~\ref{fig:regression_lollipop}, we see that the conservative and extremely conservative subgroups can be simulated better with the other LLMs instead of OLMo-32B.}
    \label{fig:regression_lollipop_tvd}
\end{figure*}

\textbf{Simulation results based on TVD. }Figure~\ref{fig:regression_lollipop_tvd} shows the same OLS analysis as Figure~\ref{fig:regression_lollipop}, but for weighted TVD instead of weighted F1. Since lower TVD is better, positive coefficients in the figure indicate a variant that performs worse than this reference (higher TVD), while negative coefficients indicate a variant that performs better. As with F1, we use the best performing variant for each design choice as the reference category; this changes which model and response handling approach is used as the baseline: for TVD, Olmo-32B is the reference LLM (rather than Qwen-30B) and free-text with an LLM judge is the reference response handling approach (rather than JSON parsing). The shift in reference categories further illustrates the \textbf{Single Best Metric Fallacy} — the "best" configuration depends on which metric is used to evaluate it. 

The persona construction results are consistent across both metrics: personas with attitudinal variables outperform demographics-only personas for both F1 and TVD, and across nearly all ideology subgroups. This is the one finding of the case study that holds up regardless of which metric is used.

The picture is less consistent for LLM selection. Like the F1 results, we see that the conservative and extremely conservative respondents are not simulated as well by the best-performing LLM. With TVD, we see that OLMo-7B also improves predictions. 

Interestingly, the adjusted $R^2$ scores are lower for the moderate and slightly conservative subgroups, even compared to the F1 regression model. This could indicate that the TVD alignment for these subgroups might be impacted by other sources of variation besides the design choices.

\begin{figure}
    \centering
    \includegraphics[width=0.72\linewidth]{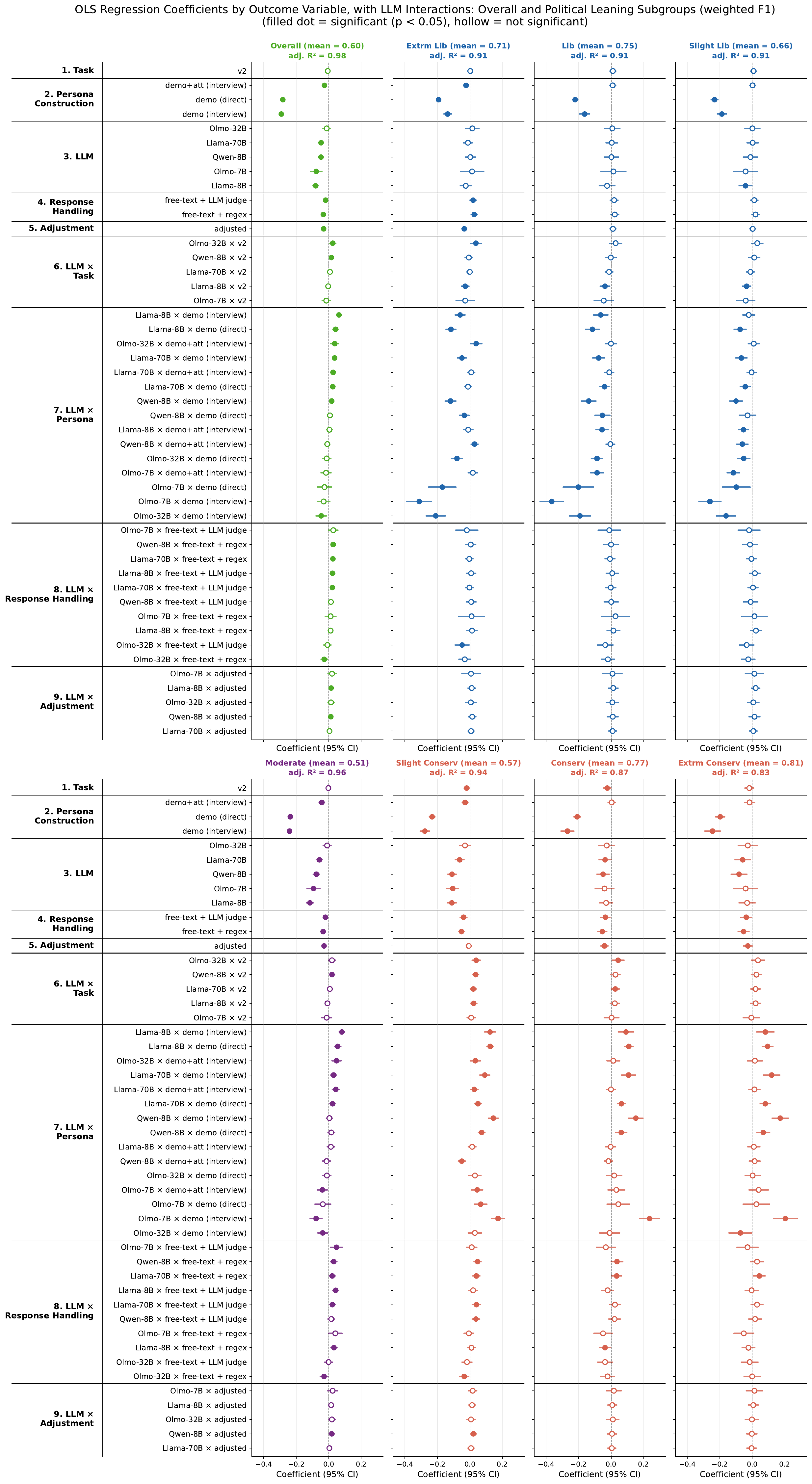}
    \caption{\textbf{Predicting F1 scores based on LLM selection, other simulation design choices, and their interaction.}}
    \label{fig:interaction_F1}
\end{figure}

\begin{figure}
    \centering
    \includegraphics[width=0.72\linewidth]{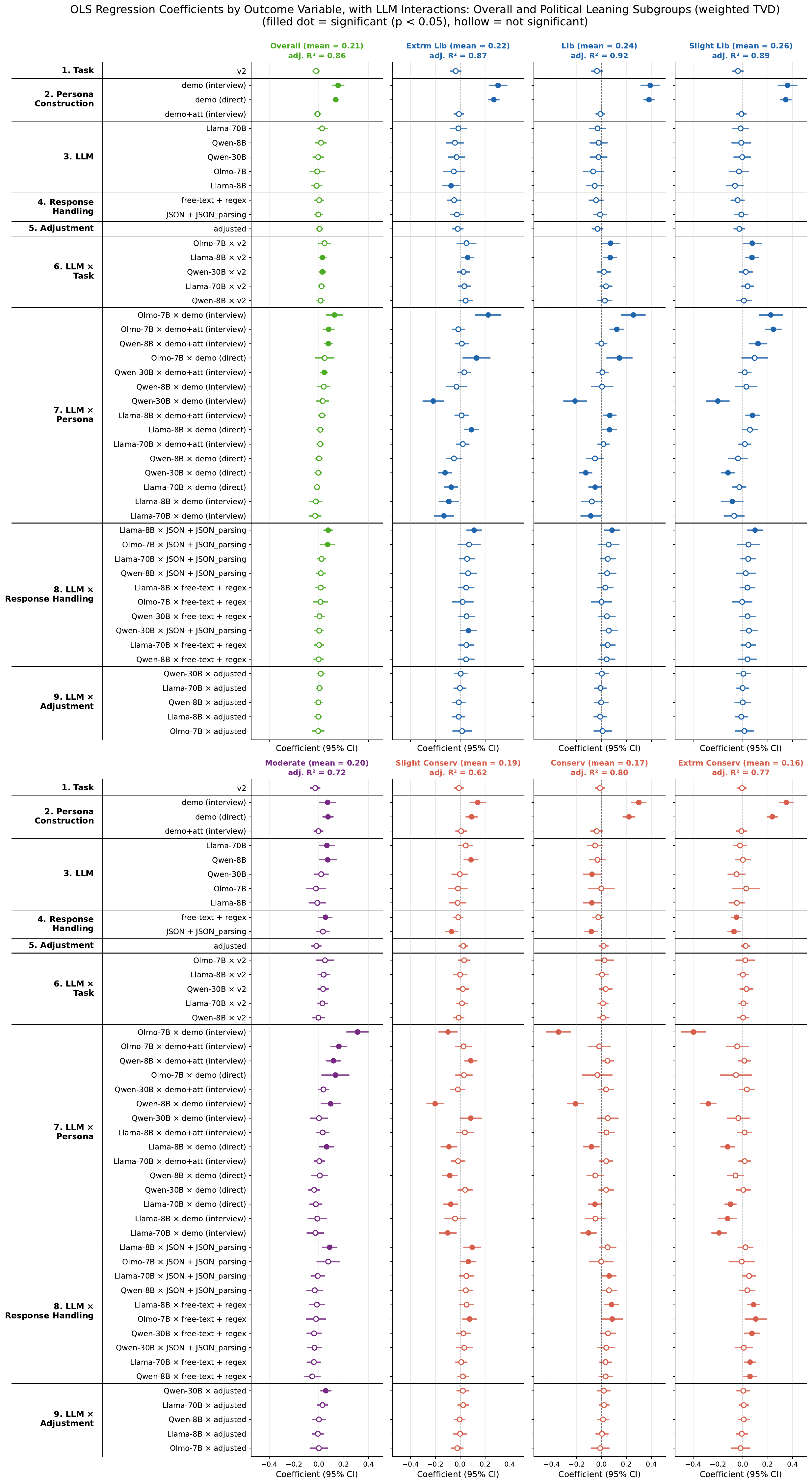}
    \caption{\textbf{Predicting TVD scores based on LLM selection, other simulation design choices, and their interaction.}}
    \label{fig:interaction_TVD}
\end{figure}

\textbf{Simulation Results with Interaction Variables. }Our regression model assesses to what extent the simulation design choices and inherent LLM behavior impacts the variation in the simulation performance. While the adjusted $R^2$ for the overall simulation is high, it is lower for different ideological subgroups. To assess if further factors could explain the variance in performance, we add interaction variables between the LLMs and the design choices. The results for performance measured with F1 scores and TVD can be found in Figures~\ref{fig:interaction_F1} and~\ref{fig:interaction_TVD}, respectively. 

We find that adding the interaction variables boosts the adjusted $R^2$ substantially, indicating that certain LLMs and certain design choices impact simulation performance, systematically, e.g., the interview-style personas and OLMo-7B. The patterns found in the regression without interaction variables still hold.

\newpage

\subsection{Guideline and Checklist for Documenting LLM-powered Survey Responses}\label{app:checklist}

To ensure better documentation and disclosure of LLM-generated survey responses, we create a  checklist that guides designers of simulated surveys. The checklist builds on the AAPOR reporting guidelines~\citep{rothschild2026responsible} and GuideLLM~\citep{feuerriegel2026reporting}. However, it  includes a deeper focus on LLM survey simulations. Based on the errors and evaluation fallacies in the TS2E, we include questions on how the target population is defined, how personas are constructed, and how the LLM-generated responses are evaluated and against what reference data. Our checklist can be combined with some of the fields in GuideLLM, specifically D1 (Data inputs \& privacy), F1 (Reproducibility), and G1(Competing interests) for full documentation. A markdown and word template for the checklist can be found in our code repository: \url{https://github.com/dess-mannheim/ts2e_anes_simulation}

Items in the checklist refer to stages across the lifecycle of LLM-generated survey responses, with a short explanation. To illustrate and exemplify items in the checklist, we use a complementary case study: predicting the 2024 European Parliament elections~\citep{vonderheyde2025uniteddiversitycontextualbiases}. 

%
%
%
%

\newlength{\colPh}   \newlength{\colCat} \newlength{\colSub}
\newlength{\colItem} \newlength{\colEx}

\setlength{\colPh}{1.85cm}
\setlength{\colCat}{2.35cm}
\setlength{\colSub}{2.65cm}
\setlength{\colItem}{4.45cm}
\setlength{\colEx}{4.20cm}


\newlength{\colSubItem}
\newlength{\colCatItem}
\makeatletter\newcommand{\ts@computespans}{%
  \setlength{\colSubItem}{\colSub}%
  \addtolength{\colSubItem}{\colItem}%
  \addtolength{\colSubItem}{2\tabcolsep}%
  \addtolength{\colSubItem}{\arrayrulewidth}%
  \setlength{\colCatItem}{\colCat}%
  \addtolength{\colCatItem}{\colSubItem}%
  \addtolength{\colCatItem}{2\tabcolsep}%
  \addtolength{\colCatItem}{\arrayrulewidth}}
\makeatother

\newcommand{\PH}[1]{\textbf{#1}}                      
\newcommand{\CAT}[2]{\textbf{#1}\newline\textit{#2}}  
\newcommand{\CATP}[1]{\textbf{#1}}                    
\newcommand{\IT}[2]{\textbf{#1}\newline #2}           
\newcommand{\ITspan}[2]{
  \multicolumn{2}{>{\RaggedRight\arraybackslash}p{\colSubItem}|}{\IT{#1}{#2}}}
\newcommand{\ITspanCat}[2]{
  \multicolumn{3}{>{\RaggedRight\arraybackslash}p{\colCatItem}|}{\IT{#1}{#2}}}
\newcommand{\TODOnote}[1]{\textcolor{red}{[#1]}}

\begingroup
\footnotesize
\setlength{\tabcolsep}{3pt}
\setlength{\arrayrulewidth}{0.4pt}
\makeatletter\ts@computespans\makeatother
\setlength{\LTcapwidth}{\textwidth}
\renewcommand{\arraystretch}{1.15}
\arrayrulecolor{black!35}

\begin{longtable}{|
  >{\RaggedRight\arraybackslash}p{\colPh}|
  >{\RaggedRight\arraybackslash}p{\colCat}|
  >{\RaggedRight\arraybackslash}p{\colSub}|
  >{\RaggedRight\arraybackslash}p{\colItem}|
  >{\RaggedRight\arraybackslash}p{\colEx}|}

\caption{TS2E documentation checklist: items to document
across the lifecycle of LLM-generated survey responses, with a short
explanation and a worked example for each.}
\label{tab:ts2e-checklist}\\

\hline
\rowcolor{black!8}
\textbf{Phase} &
\multicolumn{3}{>{\RaggedRight\arraybackslash}p{\colCatItem}|}{%
  \cellcolor{black!8}\textbf{Category / Item and explanation}} &
\cellcolor{black!8}\textbf{Example} \\
\hline
\endfirsthead

\multicolumn{5}{@{}l}{\textit{Table \ref{tab:ts2e-checklist} continued from
previous page}}\\
\hline
\rowcolor{black!8}
\textbf{Phase} &
\multicolumn{3}{>{\RaggedRight\arraybackslash}p{\colCatItem}|}{%
  \cellcolor{black!8}\textbf{Category / Item and explanation}} &
\cellcolor{black!8}\textbf{Example} \\
\hline
\endhead

\hline
\multicolumn{5}{r@{}}{\textit{continued on next page}}\\
\endfoot

\hline
\endlastfoot

\PH{P.\newline Prelimi\-naries}
  & \IT{P1. Goal of simulation}{How will the synthetic responses be used?}
  & \IT{P1.1 Pretesting}{The synthetic responses are used only to improve
      an instrument later fielded to humans}
  & n/a \\
\cline{3-5}

  &
  & \IT{P1.2 Synthetic Responses as Data}{The synthetic responses are used
      for research (scientific and/or commercial) or decision-making}
  & \IT{P1.2.1 Purpose of Study}{The substantive research goal and whether
      the aim is descriptive or relational.}
  & Predicting the results of the 2024 European parliament elections. \\
\cline{4-5}

  & &
  & \IT{P1.2.2 Degree of Substitution}{full replacement, a hybrid mix (e.g.,
      imputation), or augmentation of a human sample.}
  & Full substitution based on sample, data, and weights from an existing
    survey. \\
\cline{4-5}

  & &
  & \IT{P1.2.3 High-Stakes Case?}{Whether errors could drive consequential
      real-world decisions; high stakes raise the required rigor and
      disclosure.}
  & Experimental, not used for official polls or campaigns~-- high-stakes in
    theory, low-stakes in practice. \\
\cline{2-5}

  & \ITspanCat{P2 Hypotheses or Research Questions}{That are being tested
      through the simulation}
  & Can LLMs predict the aggregate results of future elections? How does
    LLMs' predictive performance differ across countries, across prompt
    languages, and depending on the information provided in the prompt? \\
\cline{2-5}

  & \ITspanCat{P3 Construct}{The underlying concept the survey measures,
      decomposed into facets if multidimensional, with the estimand stated
      precisely.}
  & Vote choice in the 2024 European Parliament elections (national party
    vote shares). \\
\cline{2-5}

  & \ITspanCat{P4 Target Population and Relevant Subgroups}{Whose responses
      the simulation should generalize to, plus any subgroups needing
      separate estimates or comparison, as well as the temporal content to
      which these findings apply.}
  & EU citizens eligible to vote in the 2024 elections, split by country of
    residence. \\
\hline

\PH{Before Fielding (Pre-data collection)}
  & \CATP{Q. Questionnaire}
  & \ITspan{Q1.1 Exact Question Phrasing(s)}{The verbatim wording of each
      item shown to the LLM, including instructions; reuse should cite
      source.}
  & ``Will you vote in the 2024 elections to the European Parliament and if
    so, for which party?'' \\
\cline{3-5}

  &
  & \ITspan{Q1.2 Questionnaire Presentation}{How items are organized and
      delivered~-- one at a time, as a battery, or all in one prompt, other.}
  & Only one item. Single-turn requests: LLMs are re-initiated for each
    persona/question so later responses aren't conditioned on earlier
    ones. \\
\cline{3-5}

  &
  & \ITspan{Q1.3 Question Variations}{Any alternative phrasings used for
      robustness or as experimental conditions}
  & Experimental translations to native language for France, Germany, Poland,
    Slovakia, Sweden. \\
\cline{3-5}

  &
  & \ITspan{Q1.4 Answer Option Variations}{Alternative response-option sets,
      e.g., scale length, labels, ordering, and whether ``don't
      know''/``refused'' are offered.}
  & No pre-specified answer options (free-text response with specific
    instructions) \\
\cline{3-5}

  &
  & \ITspan{Q1.5 Rationale for Questionnaire}{Justification for the item
      choices and any pretesting evidence (cognitive testing, pilots, prior
      validation) that they capture the construct.}
  & The survey from which the sample is borrowed does not provide a concrete
    voting intention question. The chosen wording has been adapted from
    similar earlier studies (e.g., ~\cite{argyle2023out,vonderheyde2025uniteddiversitycontextualbiases}) \\
\cline{2-5}

  & \IT{S. Simulation Units}{The unit of observation in the simulation}
  & \ITspan{S1 Unit Description}{The list of entities that can be simulated}
  & Individual survey respondents \\
\cline{3-5}

  &
  & \ITspan{S2 Sample Size}{The number of entities simulated and how it was
      set}
  & 25.916 (ca.\ 1000 per country; ca.\ 500 for Cyprus, Luxembourg, Malta,
    respectively, ca.\ 1500 for Germany) from the Eurobarometer survey sample
    (EB 99.4). \\
\cline{3-5}

  &
  & \ITspan{S3 Sampling Procedure}{How units were selected from the frame,
      with any stratification and the random seed. Also state if the sampling
      is probabilistics, i.e., every frame member had a known, non-zero
      selection chance.}
  & Stratified, multi-stage probability samples. \\
\cline{2-5}

  & \IT{PC. Persona}{The prompt-based persona used to steer LLMs during
      inference}
  & \ITspan{PC1 Format}{The grammatical and structural form the persona
      takes in the prompt.}
  & Second-person declarative: ``You are a voting-eligible citizen of an
    EU-member state in [country]''. \\
\cline{3-5}

  &
  & \ITspan{PC2 Persona Characteristics}{Which attributes compose the
      persona and where their values come from}
  & Country of residence, age, gender, education, class, employment status,
    urbanity; variation: plus political interest, ideology, attitude towards
    EU integration, trust in EU. \\
\cline{3-5}

  &
  & \ITspan{PC3 Rationale for Persona Content and Format}{Justification for
      the attribute set and format, ideally with evidence that they improve
      fidelity.}
  & No evidence was found for better performance of first- vs.\ second-person
    declaratives. The attributes have been identified as determinants of
    voting behavior in EU elections~\citep{braun2022issues,ford2020changing,giebler2015contrasting}.\\
\cline{2-5}

  & \CATP{PR. Prompts}
  & \ITspan{PR1 (C1 in GUIDE-LLM) Exact Prompt}{The verbatim text of the
      full prompt, including any in-context examples, exactly as sent to the
      model.}
  & Available in supplement I of~\citep{vonderheyde2025uniteddiversitycontextualbiases}\\
\cline{3-5}

  &
  & \ITspan{PR2 System vs.\ User Prompt}{Which prompt channel carries which
      content, and whether both are used}
  & No system prompt; persona and question in the user prompt. \\
\cline{3-5}

  &
  & \ITspan{PR3 Specialization}{Task-specific augmentations, e.g., few-shot
      examples, retrieved background, or injected temporal context.}
  & Temporal \& political context: ``The year is 2024. You are a
    voting-eligible EU citizen living in [country]'' initiates the prompt,
    followed by a list of parties competing in the respective country
    (randomized order). Instructions are given to make the best possible
    prediction based on the information given (to avoid refusals) and to give
    a short response, i.e.\ ``No'' or the name of the chosen party (to
    optimize matching and avoid unnecessarily long or cut-off answers). \\
\cline{3-5}

  &
  & \ITspan{PR4 Reasoning}{Whether and how the model is prompted to reason
      before answering, and whether the trace is kept.}
  & No reasoning requested. \\
\hline

\PH{Fielding (Data collection)}
  & \ITspanCat{NR. Number of Runs}{How many independent generations per
      unit.}
  & 1 run per persona and experimental setting (prompt language, prompt
    content) \\
\cline{2-5}

  & \CATP{LLM(s)}
  & \CATP{LLM1 Model(s) Chosen}
  & \IT{Link}{A precise pointer to the exact model and version, e.g., API
      string, model card, or checkpoint hash.}
  & GPT-4-Turbo (version 2024-04-09): \texttt{gpt-4-\allowbreak turbo-\allowbreak 2024-04-09};
    \texttt{meta-llama/\allowbreak Llama-3.1-\allowbreak 8B-Instruct};
    \texttt{mistralai/\allowbreak Mistral-7B-\allowbreak Instruct-v0.3} \\
\cline{4-5}

  & &
  & \IT{Rationale}{The reasons for choosing the specific LLM(s), e.g.\
      performance aspects, privacy or reproducibility aspects, etc.}
  & At the time of data collection, GPT-4-Turbo had the most recent training
    data corpus of all GPT models and was supposed to have better
    multilingual capacities, be better at solving complex instructions, and
    less likely to hallucinate. Its performance in predicting public opinion
    was shown to be better when adding information beyond demographics~\citep{lee2023can}, and in different languages~\citep{wang2024not}. Llama is
    optimized for multilingual dialogue use cases and supposed to be
    comparable to GPT but superior to other open-source LLMs. Mistral was
    chosen for its robust performance in handling a wide range of tasks,
    including those requiring content reasoning and creative writing, which
    was shown to complement and in some cases even surpass the strengths of
    Llama models \\
\cline{3-5}

  &
  & \IT{LLM2 Training Data}{Information about the LLM's training data
      (pre-training, post-training)}
  & \IT{Knowledge Cut-off}{The date past which the model has no training
      knowledge}
  & December 2022 (Mistral); December 2023 (GPT, Llama) \\
\cline{4-5}

  & &
  & \IT{Training Data Availability}{How much is publicly known about the
      training corpus and its curation.}
  & GPT: proprietary model, no information on training data. Llama, Mistral:
    open-weight models with limited information on training data. \\
\cline{3-5}

  &
  & \CATP{LLM3 Model Details}
  & \IT{Open vs.\ Proprietary}{Whether the model is open or a proprietary
      system accessed through a vendor.}
  & GPT: proprietary. Llama: open (special community license). Mistral: open
    (modified MIT license). \\
\cline{4-5}

  & &
  & \IT{Instruct/Aligned vs.\ Base}{Whether the model is
      instruction-tuned/aligned or a base model}
  & Instruction-tuned models. \\
\cline{4-5}

  & &
  & \IT{Language Capabilities}{Languages the model handles well}
  & GPT-4-Turbo is supposed to have better multilingual capacities than
    predecessors. Llama is optimized for multilingual dialogue use cases and
    supposed to be comparable to GPT but superior to other open source LLMs.
    Mistral is a European LLM trained on multilingual data and has reported
    strong performance for European languages. \\
\cline{4-5}

  & &
  & \IT{Temperature}{The sampling-temperature setting controls randomness and
      the variance of the LLM outputs, and therefore, of the simulated
      distribution.}
  & \texttt{temperature = 0.9} to retain response variance. \\
\cline{4-5}

  & &
  & \IT{Access Mode}{How the model was reached, e.g., locally (self-hosted),
      direct API, hosted interface, or embedded in a survey platform.}
  & GPT: via the REST API; Llama, Mistral: hosted locally. \\
\cline{4-5}

  & &
  & \IT{RAG-enabled}{Whether retrieval supplemented the model with external
      documents at inference, and the corpus used.}
  & n/a \\
\cline{4-5}

  & &
  & \IT{Custom fine-tuning or steering}{Any customization beyond prompting,
      e.g., fine-tuning, adapters, or activation steering and the data used.}
  & n/a \\
\cline{2-5}

  & \CATP{RG. Response Generation}
  & \ITspan{RG1 Output Mode}{The form the response takes, e.g., free text,
      structured JSON, a forced token, or log-probabilities over options.}
  & Free text. \\
\cline{3-5}

  &
  & \ITspan{RG2 Decoding Parameters}{Decoding settings beyond
      temperature~-- \texttt{max\_tokens}, \texttt{top\_p}, penalties, stop
      sequences, seed}
  & \texttt{max\_tokens = 40}, \texttt{seed = 20240528} \\
\hline

\PH{After Fielding (Post-data collection)}
  & \CATP{RP. Response Processing}
  & \ITspan{RP1 LLM-as-judge}{Whether an LLM interprets or classifies the
      raw outputs (e.g., mapping free text to a category)}
  & No LLM involved in response processing. \\
\cline{3-5}

  &
  & \ITspan{RP2 Other Automatic Classifier}{Any non-LLM processing~-- regex,
      exact/fuzzy matching, dictionaries, or trained classifiers~-- and its
      failure modes.}
  & Regex matching party names and established acronyms and synonyms, as well
    as keywords for non-voters and invalid votes. \\
\cline{2-5}

  & \CATP{RW. Respondent Weighting}
  & \ITspan{RW1 Source of Adjustment Weights}{Where weights come from, e.g.,
      reused from the human survey, newly built, or none.}
  & Reused Eurobarometer poststratification (nonresponse) weights (W87: EU28
    minus UK) \\
\cline{3-5}

  &
  & \ITspan{RW2 Weighting Approach}{The weighting method and adjustment
      variables}
  & Per-country procedure based on national survey research institutes or
    EUROSTAT universe descriptions, using marginal and intercellular
    weighting on (minimum) sex, age, region NUTS II and size of locality. \\
\cline{2-5}

  & \CATP{E. Evaluation}
  & \ITspan{E1 Reference Responses}{The `ground truth' compared against, and
      its potential limits, i.e., it may sit in training data or itself be
      error-laden.}
  & Official per-country EU election results (turnout and party vote
    shares). \\
\cline{3-5}

  &
  & \ITspan{E2 Evaluation Metric}{The metric(s) and the level (individual,
      distribution, relationship)}
  & Aggregate estimates: (Average) absolute difference (party vote shares per
    country and turnout). Match of winning parties, party rank orderings. \\
\cline{3-5}

  &
  & \ITspan{E3 Rationale for Metric and Reference}{Justification that the
      metric and reference fit the estimand and stakes, and what they do not
      capture.}
  & Absolute differences to observed outcomes are reported because point
    estimates were the goal. No variance or confidence intervals for the
    estimates are reported because doing so would imply that the primary
    source of error stems solely from the sampling of observations, but
    additional sources of error can arise from biases inherent in LLMs'
    data-generating process. Alternative measures (match of winning party,
    rank order) are reported to allow for more lenience. No individual-level
    responses of intended/reported vote choice are available for the sample
    at the time of data collection. \\
\cline{3-5}

  &
  & \ITspan{E4 Subgroup Analysis}{Whether alignment is reported by
      subgroup.}
  & Separate party vote shares per country-party. \\
\cline{3-5}

  &
  & \ITspan{E5 Inter-run Variance}{Whether variability across repeated runs
      is reported.}
  & No repeated runs. \\

\end{longtable}
\endgroup

\end{document}